\documentclass[10pt,a4paper]{article}
\usepackage{jcappub}

\usepackage{amsmath,amsfonts}
\usepackage{epsfig,multicol,bbm}

\newcommand{\ud}{\mathrm{d}}
\newcommand{\ve}{\varepsilon}

\newtheorem{definition}{Definition}[section]
\newtheorem{theorem}{Theorem}[section]

\title{Uniting cosmological epochs through the twister solution in cosmology
with non-minimal coupling}

\author[a]{Orest Hrycyna}
\author[b,c]{and Marek Szyd{\l}owski}
\affiliation[a]{Department of Theoretical Physics, The John Paul II Catholic
University of Lublin, Al. Rac{\l}awickie 14, 20-950 Lublin, Poland}
\affiliation[b]{Astronomical Observatory, Jagiellonian University, Orla 171,
30-244 Krak{\'o}w, Poland}
\affiliation[c]{Mark Kac Complex Systems Research Centre, Jagiellonian
University, Reymonta 4, 30-059 Krak{\'o}w, Poland}

\emailAdd{hrycyna@kul.lublin.pl}
\emailAdd{uoszydlo@cyf-kr.edu.pl}

\date{\today}

\abstract{
We investigate dynamics of a flat FRW cosmological model with a barotropic
matter and a non-minimally coupled scalar field (both canonical and phantom). In
our approach we do not assume any specific form of a potential function for the
scalar field and we are looking for generic scenarios of evolution. We show that
dynamics of universe can be reduced to a 3-dimensional dynamical system. We have
found the set of fixed points and established their character. These critical
points represent all important epochs in evolution of the universe : (a) a
finite scale factor singularity, (b) an inflation (rapid-roll and slow-roll),
(c) a radiation domination, (d) a matter domination and (e) a quintessence era.
We have shown that the inflation, the radiation and matter domination epochs are
transient ones and last for a finite amount of time. The existence of the radiation
domination epoch is purely the effect of a non-minimal coupling constant. We show
the existence of a twister type solution wandering between all these critical
points.
}

\arxivnumber{1008.1432}

\begin{document}
\maketitle
\section{Introduction}

In modern cosmology a scalar field $\phi$ plays a very special role. The
discovery of accelerated cosmic expansion \cite{Riess:1998cb,
Perlmutter:1998np} gave a motivation to study dynamics of dark energy models
(see \cite{Copeland:2006wr} for review). In the context of the quintessence idea
\cite{Ratra:1987rm, Wetterich:1987fm} the simplest dynamical models involving
the scalar field $\phi$ with a potential function $V(\phi)$ are used to model a
time dependent equation of state parameter $w_{\phi}$. While the simplest
candidate for dark energy seems to be a positive cosmological constant, the
$\Lambda$CDM model is favoured by observational data \cite{Kurek:2007tb,
Kurek:2007gr, Szydlowski:2006ay}, such a explanation of cosmic acceleration
suffers from the fine tuning problem \cite{Weinberg:1988cp} and the coincidence
problem \cite{Zlatev:1998tr}. In order to alleviate those problems many
alternatives have been proposed like phantom dark energy \cite{Caldwell:2003vq,
Dabrowski:2003jm} or extended quintessence \cite{Faraoni:2006ik, Faraoni:2000gx,
Hrycyna:2007mq, Hrycyna:2007gd}.

If we are going to generalise the scalar field cosmology minimally coupled with
gravity, then inclusion of the non-minimal coupling term of type $-\xi
R \phi^{2}$ \cite{Chernikov:1968zm, Callan:1970ze, Birrell:1979ip} seems to be
natural and the simplest generalisation of the Lagrangian for scalar field
dynamics in the background of cosmological models with maximal symmetry of
space-like slices. Of course the value of this additional parameter should be
estimated from observational data \cite{Luo:2005ra, Jankiewicz:2005tm,
Szydlowski:2008zza} or given from some theoretical arguments
\cite{Faraoni:2000gx}. The nonzero $\xi$ arises from quantum corrections
\cite{Birrell:1984ix} and it is required by the renormalization
\cite{Callan:1970ze}. While the simplest minimally coupled scalar field with a
quadratic potential function has strong motivations in observations
\cite{Komatsu:2008hk, Komatsu:2010fb} its generalisations with a non-minimal
coupling term was studied \cite{Park:2008hz} in the context of origin of the
canonical inflaton itself.

In this paper we investigate the dynamical evolution of scalar field
cosmological models with a non-vanishing coupling constant between a scalar
field and gravity. The role of the non-minimal coupling in evolution of the
universe in the context of inflation and quintessence was studied
previously by many authors \cite{Spokoiny:1984bd,
Salopek:1988qh, Fakir:1992cg, Belinsky:1985zd,
Barvinsky:1994hx, Barvinsky:1998rn, Barvinsky:2008ia, Setare:2008mb,
Setare:2008pc,Hrycyna:2008gk, Uzan:1999ch, Chiba:1999wt, Amendola:1999qq,
Perrotta:1999am, Holden:1999hm, Bartolo:1999sq, Boisseau:2000pr,
Gannouji:2006jm, Carloni:2007eu, Bezrukov:2007ep, Nozari:2007eq,
Kamenshchik:1995ib} and in connection with the development of the Standard Model
with a non-minimally coupled Higgs field \cite{DeSimone:2008ei, Bezrukov:2008ej,
Barvinsky:2009fy, Clark:2009dc}. The dynamical systems methods are used in investigating
of evolutional paths of cosmological models which dynamics is parameterised by
the energetic variables very useful in this context \cite{Copeland:1997et}.

We are adopting dynamical systems methods in studying the evolution of the
cosmological model and its dynamics can be visualised in the phase space which
is a geometric framework for its exploration. Moreover one can investigate all
evolutional paths of the system under consideration for all admissible initial
conditions. Therefore one should ask whether different models with a desired
property are typical (generic) in the class of all possible models. In our
opinion physically interesting cosmological models should be generic in the
sense that they do not depend on the special choice of initial conditions which
should be determined from quantum models.

To keep generality of our considerations we do not assume any specific form of
the potential function of a canonical or phantom scalar field. It will be
demonstrated that the parameter of non-minimal coupling $\xi$ plays the crucial
role during the cosmological evolution. We will show the emergence of a new
phase space structure organised through critical points and trajectories. For
completeness we also include the model with the barotropic matter with the
constant equation of state parameter $w_{m}$.

We demonstrate that, in principle, the phase space structure at a finite domain
is determined by five critical points corresponding to important events during
the cosmic evolution, namely, the singularity (of a finite scale factor type),
inflation, radiation and matter dominated epochs and finally the accelerated
expansion era. In our previous paper we introduced notion of the twister
solutions the solutions linking the subsequent cosmological epochs
\cite{Hrycyna:2009zj}. In the present paper we generalise this notion without
assuming any form of the potential function of the scalar field. The evolutional
scenarios investigated in this paper are obvious
only if the non-minimal coupling is different from minimal $(\xi=0)$ and
conformal $(\xi=1/6)$ coupling value. In this sense we study a unique type of
evolution.

\section{The model}

In the model under consideration we assume the spatially flat
Friedmann-Robertson-Walker (FRW) universe filled with the non-minimally coupled
scalar field and barotropic fluid with the equation of the state coefficient
$w_{m}$. The action assumes following form
\begin{equation}
S = \frac{1}{2}\int \ud^{4}x \sqrt{-g} \Bigg(\frac{1}{\kappa^{2}}R - \ve
\Big(g^{\mu\nu}\partial_{\mu}\phi\partial_{\nu}\phi + \xi R \phi^{2}\Big) -
2U(\phi) \Bigg) + S_{m},
\end{equation}
where $\kappa^{2}=8\pi G$, $\ve = +1,-1$ corresponds to canonical and phantom
scalar field, respectively, the metric signature is $(-,+,+,+)$,
$R=6\Big(\frac{\ddot{a}}{a}+\frac{\dot{a}^{2}}{a^{2}}\Big)$ is the Ricci scalar,
$a$ is the scale factor and a dot denotes differentiation with respect to the
cosmological time and $U(\phi)$ is the scalar field potential function. $S_{m}$
is the action for the barotropic matter part.

The dynamical equation for the scalar field we can obtain from the variation
$\delta S/\delta \phi = 0$
\begin{equation}
\ddot{\phi} + 3 H \dot{\phi} + \xi R \phi + \ve U'(\phi) =0,
\end{equation}
and energy conservation condition from the variation $\delta S/\delta
g^{\mu\nu}=0$
\begin{equation}
\mathcal{E}= \ve \frac{1}{2}\dot{\phi}^{2} + \ve3\xi H^{2}\phi^{2} + \ve3\xi H
(\phi^{2})\dot{} + U(\phi) + \rho_{m} - \frac{3}{\kappa^{2}}H^{2}.
\end{equation}
Then conservation conditions read
\begin{eqnarray}
\frac{3}{\kappa^{2}}H^{2} & = & \rho_{\phi} + \rho_{m}, \\
\dot{H} & = & -\frac{\kappa^{2}}{2}\Big[(\rho_{\phi}+p_{\phi}) +
\rho_{m}(1+w_{m})\Big]
\end{eqnarray}
where the energy density and the pressure of the scalar field are
\begin{eqnarray}
\rho_{\phi} & = & \ve\frac{1}{2}\dot{\phi}^{2}+U(\phi)+\ve3\xi H^{2}\phi^{2} +
\ve 3\xi H (\phi^{2})\dot{},\\
p_{\phi} & = & \ve\frac{1}{2}(1-4\xi)\dot{\phi}^{2} - U(\phi) + \ve\xi
H(\phi^{2})\dot{} - \ve2\xi(1-6\xi)\dot{H}\phi^{2} -
\ve3\xi(1-8\xi)H^{2}\phi^{2} + 2\xi\phi U'(\phi).
\end{eqnarray}

Note that, when the non-minimal coupling is present, the energy density
$\rho_{\phi}$ and the pressure $p_{\phi}$ of the scalar field can be defined in
several possible inequivalent ways. This corresponds to different ways
of writing the field equations. In the case adopted here the energy momentum
tensor of the scalar field is covariantly conserved, which may not be true for
other choices of $\rho_{\phi}$ and $p_{\phi}$
\cite{Faraoni:2000wk,Faraoni:2004pi}. For example, the redefinition of
gravitational constant $\kappa_{\text{eff}}^{-2}=\kappa^{-2}-\ve\xi\phi^{2}$ makes it
time dependent. The effective gravitational constant can diverge for a critical
value of the scalar field $\phi_{c}=\pm(\ve\kappa^{2}\xi)^{-1/2}$. Though the FRW
model remains regular at this point, the model is unstable with respect to
arbitrary small anisotropic and inhomogeneous perturbations which become
infinite there. This results in the formation of a strong curvature singularity
prohibiting a transition to the region $\kappa_{\text{eff}}^{2}<0$
\cite{Starobinsky:1981}.

In what follows we introduce the energy phase space variables
\begin{equation}
\label{eq:envar}
x\equiv \frac{\kappa \dot{\phi}}{\sqrt{6}H}, \quad
y\equiv\frac{\kappa\sqrt{U(\phi)}}{\sqrt{3}H}, \quad
z\equiv\frac{\kappa}{\sqrt{6}}\phi,
\end{equation}
which are suggested by the conservation condition
\begin{equation}
\frac{\kappa^{2}}{3H^{2}}\rho_{\phi} + \frac{\kappa^{2}}{3H^{2}}\rho_{m} =
\Omega_{\phi} + \Omega_{m} = 1
\end{equation}
or in terms of the newly introduced variables
\begin{equation}
\Omega_{\phi} = y^{2} + \ve\Big[(1-6\xi)x^{2}+6\xi(x+z)^{2}\Big] = 1
-\Omega_{m}.
\end{equation}

The acceleration equation can be rewritten to the form
\begin{equation}
\label{eq:accel}
\dot{H} = -\frac{\kappa^{2}}{2}\Big(\rho_{\rm{eff}}+p_{\rm{eff}}\Big) =
-\frac{3}{2}H^{2}(1+w_{\rm{eff}})
\end{equation}
where the effective equation of the state parameter reads
\begin{eqnarray}
w_{\rm{eff}}=\frac{1}{1-\ve6\xi(1-6\xi)z^{2}}\Big[ -1 +
\ve(1-6\xi)(1-w_{m})x^{2} + \ve2\xi(1-3w_{m})(x+z)^{2} + \nonumber \\ 
 + (1+w_{m})(1-y^{2}) -\ve2\xi(1-6\xi)z^{2} - 2\xi y^{2} \lambda z\Big]
\label{eq:weff}
\end{eqnarray}
where $\lambda = -\frac{\sqrt{6}}{\kappa}\frac{1}{U(\phi)}\frac{\ud U(\phi)}
{\ud\phi}$.

The dynamical system describing the investigated models is in the following
form \cite{Szydlowski:2008in, Hrycyna:2009zj}
\begin{subequations}
\label{eq:dynsys1}
\begin{eqnarray}
\label{eq:dynsys1a}
x' & = & -(x-\ve\frac{1}{2}\lambda
y^{2})\Big[1-\ve6\xi(1-6\xi)z^{2}\Big] +
\frac{3}{2}\left(x+6\xi z\right) 
\bigg[ -\frac{4}{3} - 2\xi \lambda y^{2} z + \nonumber \\
 & &   + \ve(1-6\xi)(1-w_{m})x^{2}
 +\ve2\xi(1-3w_{m})\left(x+z\right)^{2} + (1+w_{m})(1-y^{2}) \bigg],\\
 \label{eq:dynsys1b}
 y' & = & y\left(2-\frac{1}{2}\lambda x\right)
 \Big[1-\ve6\xi(1-6\xi)z^{2}\Big]
 +\frac{3}{2} y \bigg[ -\frac{4}{3} - 2\xi \lambda y^{2} z + \nonumber
 \\
 & & + \ve(1-6\xi)(1-w_{m})x^{2} +
 \ve2\xi(1-3w_{m})\left(x+z\right)^{2}
 + (1+w_{m})(1-y^{2})\bigg],\\
 \label{eq:dynsys1c}
 z' & = & x \Big[1-\ve6\xi(1-6\xi)z^{2}\Big],\\
 \label{eq:dynsys1d}
 \lambda' & = & -\lambda^{2}\left(\Gamma-1\right) x
\Big[1-\ve6\xi(1-6\xi)z^{2}\Big].
\end{eqnarray}
\end{subequations}
where a prime denotes differentiation with respect to time $\tau$ defined as
\begin{equation}
\frac{\ud}{\ud\tau} =
\bigg[1-\ve6\xi(1-6\xi)z^{2}\bigg]\frac{\ud}{\ud\ln{a}}
\label{eq:timerep1}
\end{equation}
where the expression in brackets is assumed as a positive quantity to assure
that during the evolution with $\tau>0$ the scale factor $a$ is growing, i.e.
the universe expands, and
$$
\Gamma = \frac{U''(\phi)U(\phi)}{U'(\phi)^{2}}.
$$
To investigate the dynamics of the universe described by the dynamical system
(\ref{eq:dynsys1}) we need to define an unknown function $\Gamma$, i.e. we need
to define the potential function $U(\phi)$. In the special cases of the system
with the cosmological constant or exponential potential, $U=U_{0}=\text{const.}$
or $U=U_{0}\exp{(-\lambda \phi)}$, the dynamical system (\ref{eq:dynsys1}) can
be reduced to the $3$-dimensional one due to the relation $\lambda=0$ and 
$\Gamma=0$ in the former case, and $\lambda=\text{const.}$ and $\Gamma=1$ in 
the latter case. Then dynamical system consists of three
equations (\ref{eq:dynsys1a}, \ref{eq:dynsys1b}, \ref{eq:dynsys1c}).

\begin{table*}
\begin{tabular}{|c|l|l|}
\hline\hline
parameters & $z(\lambda)$ & potential function $U(\phi)$ \\
\hline
$\alpha\ne0$, $\beta=0$, $\gamma=0$ & $\frac{\lambda}{\alpha} + {\rm const.}$ & $U_{0}\exp{\left(-\frac{\alpha}{2}\phi^{2} + {\rm const.}\phi\right)}$ \\

$\alpha=0$, $\beta\ne0$, $\gamma=0$ & $\frac{\ln{\lambda}}{\beta} + {\rm
const.}$  & $U_{0}\exp{\left(\frac{{\rm const.}}{\beta}\exp{\left(\beta \phi\right)}\right)}$ \\

$\alpha=0$, $\beta=0$, $\gamma\ne0$ & 
$-\frac{1}{\gamma\lambda} + {\rm const.}$ & 
$U_{0}\left(\gamma\phi - {\rm const.}\right)^{\frac{1}{\gamma}}$ \\

$\alpha\ne0$, $\beta\ne0$, $\gamma=0$ &
$\frac{\ln{(\alpha+\beta\lambda)}}{\beta} + {\rm const.}$ & 
$U_{0}\exp{\left(\frac{1}{\beta}\left(\alpha\phi+{\rm conts.}\exp{(\beta\phi)}\right)\right)}$ \\

$\alpha\ne0$, $\beta=0$, $\gamma\ne0$ & 
$\frac{\arctan{\left(\sqrt{\frac{\gamma}{\alpha}}\lambda\right)}}{\sqrt{\alpha\gamma}} + {\rm const.}$ &
$U_{0}\left(\cos{\left(\sqrt{\alpha\gamma}(\phi - {\rm const.})\right)}\right)^{\frac{1}{\gamma}}$ \\

$\alpha=0$, $\beta\ne0$, $\gamma\ne0$ &
 $\frac{\ln{\lambda} -  \ln{(\beta+\gamma\lambda)}}{\beta} + {\rm const.}$ & 
$U_{0}\left(\exp{\left({\rm const.}\beta\right)}+\gamma\exp{\left(\beta\phi\right)}\right)^{\frac{1}{\gamma}}$ \\

$\alpha\ne0$, $\beta\ne0$, $\gamma\ne0$ & 
$\frac{2\arctan{\left(\frac{\beta+2\gamma\lambda}{\sqrt{-\beta^{2}+4\alpha\gamma}}\right)}}{\sqrt{-\beta^{2}+4\alpha\gamma}}
+ {\rm const.}$ &
$U_{0}\exp{\left(\frac{\beta}{2\gamma}\phi\right)}\left(\cos{\left(\frac{1}{2}\sqrt{-\beta^{2}+4\alpha\gamma}(\phi-{\rm
const.})\right)}\right)^{\frac{1}{\gamma}}$ \\
\hline\hline
\end{tabular}
\caption{\label{tab:1} Different examples of potential functions for various
configurations of parameters values of the assumed form of the $\Gamma(\lambda)$
function $\Gamma(\lambda)=1-\frac{1}{\lambda^{2}}\left(\alpha+\beta\lambda+
\gamma\lambda^{2}\right)$.}
\end{table*}

There is another possibility of reduction of the system (\ref{eq:dynsys1})
from a $4$-dimensional dynamical system to a $3$-dimensional one. If we assume that
$z=z(\lambda)$ and $\Gamma=\Gamma(\lambda)$, then using (\ref{eq:dynsys1c}) and
(\ref{eq:dynsys1d}) we can find the function $z(\lambda)$ from the differential
equation
\begin{equation}
\frac{\ud z(\lambda)}{\ud \lambda} = z'(\lambda) =
-\frac{1}{\lambda^{2}\big(\Gamma(\lambda)-1\big)}
\end{equation}
which can be integrated for some given function $\Gamma(\lambda)$
\begin{equation}
\label{eq:zlambda}
z(\lambda) = - \int \frac{\ud \lambda}{\lambda^{2}\big(\Gamma(\lambda)-1\big)}.
\end{equation}
Then the dynamical system describing the investigated models is in the following
form \cite{Szydlowski:2008in, Hrycyna:2009zj}
\begin{subequations}
\label{eq:dynsys}
\begin{eqnarray}
x' & = & -(x-\ve\frac{1}{2}\lambda
y^{2})\Big[1-\ve6\xi(1-6\xi)z(\lambda)^{2}\Big] +
\frac{3}{2}\left(x+6\xi z(\lambda)\right) 
\bigg[ -\frac{4}{3} - 2\xi \lambda y^{2} z(\lambda) + \nonumber \\
 & &   + \ve(1-6\xi)(1-w_{m})x^{2}
+\ve2\xi(1-3w_{m})\left(x+z(\lambda)\right)^{2} + (1+w_{m})(1-y^{2}) \bigg],\\
y' & = & y\left(2-\frac{1}{2}\lambda x\right)
\Big[1-\ve6\xi(1-6\xi)z(\lambda)^{2}\Big]
+\frac{3}{2} y \bigg[ -\frac{4}{3} - 2\xi \lambda y^{2} z(\lambda) + \nonumber
\\
& & + \ve(1-6\xi)(1-w_{m})x^{2} + \ve2\xi(1-3w_{m})\left(x+z(\lambda)\right)^{2}
+ (1+w_{m})(1-y^{2})\bigg],\\
\lambda' & = & -\lambda^{2}\left(\Gamma(\lambda)-1\right) x
\Big[1-\ve6\xi(1-6\xi)z(\lambda)^{2}\Big].
\end{eqnarray}
\end{subequations}
where a prime denotes now differentiation with respect to time $\tau$ defined as
\begin{equation}
\frac{\ud}{\ud\tau} =
\bigg[1-\ve6\xi(1-6\xi)z(\lambda)^{2}\bigg]\frac{\ud}{\ud\ln{a}}
\label{eq:timerep}
\end{equation}
and we assume that the term in bracket is positive in the phase space during the
evolution.

Now we are able to express the acceleration equation (\ref{eq:accel}) in
terms of the energy phase space variables and time $\tau$
\begin{equation}
\frac{\ud \ln{H^{2}}}{\ud \tau} =
-3\Big[1-\ve6\xi(1-6\xi)z\big(\lambda(\tau)\big)^{2}\Big]\big(1+w_{\text{eff}}\big)
\end{equation}
which together with (\ref{eq:weff}) results in
\begin{equation}
\label{eq:H2}
\begin{array}{ccl}
\ln{\left(\frac{H}{H^{\text{ini}}}\right)^{2}} & = &
-3\int_{0}^{\tau}\Big\{1+w_{m}+ \ve(1-6\xi)(1-w_{m})x(\tau)^{2} +
\ve2\xi(1-3w_{m})\Big(x(\tau)+z\big(\lambda(\tau)\big)\Big)^{2} - \\ & & \qquad
\quad -
y(\tau)^{2}\Big(2\xi\lambda(\tau)z\big(\lambda(\tau)\big)+1+w_{m}\Big)-\ve8\xi(1-6\xi)z\big(\lambda(\tau)\big)^{2}\Big\}\ud\tau,
\end{array}
\end{equation}
where $H^{\text{ini}}$ denotes the initial value of Hubble's function at time
$\tau=0$. In what
follows we will be using this expression together with the linearised
solutions in the vicinity of every critical point to investigate the behaviour
of Hubble's function with respect to the scale factor, as well as
Hubble's radius defined as
$$
R_{H} = \frac{1}{H}.
$$

For example if the function $\Gamma(\lambda)$ is assumed in the following form
$$
\Gamma(\lambda) = 1 - \frac{1}{\lambda^{2}}\big(\alpha + \beta\lambda +
\gamma\lambda^{2}\big),
$$
then in Table~\ref{tab:1} we have gathered forms of the functions $z(\lambda)$
and corresponding potential functions $U(\phi)$ for various
configurations of values of parameters $\alpha$, $\beta$ and $\gamma$. As we see
there are various potential functions which are the most common used in the
literature of the subject. Of course this simple ansatz for the function
$\Gamma(\lambda)$ does not manage all possible potential functions. Let us
consider the following function
$$
\Gamma(\lambda)=\frac{3}{4} -
\frac{\sigma^{2}\lambda^{2}}{4\big(2\pm\sqrt{4+\sigma^{2}\lambda^{2}}\big)^{2}}
$$
as one can check from (\ref{eq:zlambda}) we receive
$$
z(\lambda)= - \frac{2\pm\sqrt{4+\sigma^{2}\lambda^{2}}}{\lambda} + \text{const.}
$$
and this example corresponds to the Higgs potential
$$
U(\phi) = U_{0}\Big((\phi-\text{const.})^{2} - \sigma^{2}\Big)^{2}.
$$
We need to stress that the discussion presented below
is not restricted to the specific potential function but is generic in the sense
that it is valid for any function $\Gamma(\lambda)$ for which the integral defined
in (\ref{eq:zlambda}) exists.

\section{Dynamics of universe with a potential}

In our investigations of dynamics of the system given by eqs.~(\ref{eq:dynsys})
we will restrict ourselves to the finite region of the phase space, i.e. we
will be interested only in the critical points in the finite domain of the
phase space. The full investigations of the dynamics requires examination of
critical points at infinity, i.e. compactification of the phase space with the
Poincar{\'e} sphere. The procedure of transforming dynamical variables into the
projective variables associated with the compactification requires that the
right hand sides of the dynamical system should be polynomial. In our case this
is not always true because of the form of function $z(\lambda)$ (see 
table~\ref{tab:1}). In what follows we present detailed discussion of character
of critical points of the system (\ref{eq:dynsys}) corresponding to different
stages of cosmological evolution (table~\ref{tab:2}).
\begin{table*}
\begin{tabular}{|l|c|c|c|c|}
\hline\hline
 & $x^{*}$  & $y^{*} $ &  $\lambda^{*}$ & $w_{\text{eff}}$ \\
 \hline
1. & $x^{*}_{1}=-6\xi z(\lambda^{*}_{1})$ & $y^{*}_{1}=0$ & $\lambda^{*}_{1} \colon
z(\lambda)^{2}=\frac{1}{\ve6\xi(1-6\xi)}$ & $\pm\infty$ \\
\hline
2a. & $x^{*}_{2a} = -6\xi z(\lambda^{*}_{2a})$ &
$(y^{*}_{2a})^{2}=\frac{4\xi}{2\xi
\lambda^{*}_{2a}z(\lambda^{*}_{2a})+(1+w_{m})}$ & $\lambda^{*}_{2a} \colon
z(\lambda)^{2} = \frac{1}{\ve6\xi(1-6\xi)}$ & $w_{m}-4\xi$ \\
2b. & $x^{*}_{2b} = 0$ & $(y^{*}_{2b})^{2} =
\frac{2\xi(1-3w_{m})}{(1-6\xi)\big(2\xi\lambda^{*}_{2b}z(\lambda^{*}_{2b})+(1+w_{m})\big)}$
& $\lambda^{*}_{2b} \colon z(\lambda)^{2} = \frac{1}{\ve6\xi(1-6\xi)}$ &
$\frac{w_{m}-2\xi}{1-6\xi}$ \\
\hline
3a. & $x^{*}_{3a}:g(x)=0$ 
\footnote{$g(x)=\ve(1-4\xi-w_{m})x^{2}+\ve4\xi(1-3w_{m})z(\lambda^{*}_{3a})x
+\frac{2\xi}{1-6\xi}(1-3w_{m})$} 
& $y^{*}_{3a}=0$ & $\lambda^{*}_{3a} \colon
z(\lambda)^{2} = \frac{1}{\ve6\xi(1-6\xi)}$ & $\frac{1}{3}$ \\
3b. & $x^{*}_{3b}=0$ & $ y^{*}_{3b}=0$ & $\lambda^{*}_{3b} \colon z(\lambda)^{2} =
\frac{1}{\ve6\xi}$ & $\frac{1}{3}$ \\
\hline
4. & $x^{*}_{4} = 0$ & $y^{*}_{4} = 0$ & $\lambda^{*}_{4} \colon z(\lambda)=0$ &
$w_{m}$ \\
\hline
5. & $x^{*}_{5} = 0$ & $(y^{*}_{5})^{2} = 1 - \ve6\xi z(\lambda^{*}_{5})^{2}$ &
$\lambda^{*}_{5} \colon \lambda z(\lambda)^{2} + 4 z(\lambda)
-\frac{\lambda}{\ve6\xi} = 0$ & $-1$ \\
\hline\hline
\end{tabular}
\makebox[\textwidth][l]{
\scriptsize $^{1} g(x)=\ve(1-4\xi-w_{m})x^{2}+\ve4\xi(1-3w_{m})z(\lambda^{*}_{3a})x
+\frac{2\xi}{1-6\xi}(1-3w_{m})$}\par
\caption{\label{tab:2} Critical points of the system under consideration.}
\end{table*}

\subsection{Finite scale factor initial singularity}

Our discussion of the dynamics of the model under consideration we begin with
the critical point located at
\begin{equation}
\label{eq:cp1}
x^{*}_{1} = -6\xi z(\lambda^{*}_{1}), \quad y^{*}_{1} = 0, \quad \lambda^{*}_{1} \colon
z(\lambda)^{2} = \frac{1}{\ve6\xi(1-6\xi)} 
\end{equation}
where the last expression means that the coordinate $\lambda^{*}_{1}$ of the
critical point is the solution to the equation $ z(\lambda)^{2} =
\frac{1}{\ve6\xi(1-6\xi)} $. This critical point represents a singularity
because the value of $w_{\rm{eff}}$ given by (\ref{eq:weff}) calculated at this
point is 
$$
w_{\rm{eff}} = \pm \infty.
$$
We need to stress that this critical point exists only if $\ve\xi(1-6\xi)>0$,
i.e. for the canonical scalar field $(\ve=+1)$ for $0<\xi<1/6$, and for the
phantom scalar field $(\ve=-1)$ for $\xi<0$ or $\xi>1/6$.

In cosmological investigations one encounters usually various types of
singularities such as: initial finite scale factor singularity
\cite{Barrow:1990td, Cannata:2008xc}, future finite scale factor singularities:
the sudden future singularities \cite{Barrow:2004xh}, the Big Brake singularity
\cite{Gorini:2003wa}, and the Big Boost singularity \cite{Barvinsky:2008rd}.

Linearised solutions in the vicinity of this critical point are
\begin{subequations}
\label{eq:sol1}
\begin{eqnarray}
x_{1}(\tau) & = & x^{*}_{1} +
\left((x^{\text{ini}}_{1}-x^{*}_{1})+2(1-3\xi)z'(\lambda^{*}_{1})(\lambda^{\text{ini}}_{1}-\lambda^{*}_{1})\right)
\exp{(l_{1}\tau)} - \nonumber \\ & & -
2(1-3\xi)z'(\lambda^{*}_{1})(\lambda^{\text{ini}}_{1}-\lambda^{*}_{1})\exp{(l_{3}\tau)},\\
y_{1}(\tau) & = & y^{\text{ini}}_{1} \exp{(l_{2}\tau)}, \\
\lambda_{1}(\tau) & = & \lambda^{*}_{1} +
(\lambda^{\text{ini}}_{1}-\lambda^{*}_{1})\exp{(l_{3}\tau)}.
\label{eq:sol1c}
\end{eqnarray}
\end{subequations}
where $$l_{1}=6\xi, \quad l_{2}=6\xi, \quad l_{3}=12\xi$$ are the eigenvalues of
the linearization matrix calculated at this critical point, $x^{\text{ini}}_{1}$,
$y^{\text{ini}}_{1}$ and $\lambda^{\text{ini}}_{1}$ are initial conditions. For positive
values of the coupling constant $\xi>0$ this critical point represents an
unstable node type critical point. For $\xi<0$ which is possible only for the
phantom scalar field $(\ve=-1)$ the critical point is of a stable node type.

Using the time reparameterization (\ref{eq:timerep})
\begin{equation}
\ud \ln{a} = \left(1-\ve6\xi(1-6\xi)z\big(\lambda(\tau)\big)^{2}\right) \ud \tau
\end{equation}
and expansion into the Taylor series around the critical point coordinate
$\lambda^{*}$ 
$$
z(\lambda) = z(\lambda^{*}) +z'(\lambda^{*})(\lambda-\lambda^{*})
$$
up to linear terms
$$
z(\lambda)^{2} = z(\lambda^{*})^{2} + 2
z(\lambda^{*})z'(\lambda^{*})(\lambda-\lambda^{*})
$$
and then together with (\ref{eq:sol1c}) we have
$$
z\left(\lambda(\tau)\right)^{2} = z(\lambda^{*}_{1})^{2} + 2
z(\lambda^{*}_{1})z'(\lambda^{*}_{1})(\lambda^{\text{ini}}_{1}-\lambda^{*}_{1})\exp{(l_{3}\tau)}.
$$
Inserting this expansion into the time reparameterization we receive
$$
\ud \ln{a} = - \ve 12 \xi (1-6\xi) z(\lambda^{*}_{1})z'(\lambda^{*}_{1})
(\lambda^{\text{ini}}_{1}-\lambda^{*}_{1}) \exp{(l_{3}\tau)} \ud \tau
$$
which can be directly integrated for $l_{3}=12\xi>0$
$$
\Delta \ln{a} = -\ve12\xi(1-6\xi)z(\lambda^{*}_{1})z'(\lambda^{*}_{1})
(\lambda^{\text{ini}}_{1}-\lambda^{*}_{1})\int_{-\infty}^{0} \exp{(l_{3}\tau)} \ud\tau
$$
(we could take also $l_{3}=12\xi<0$ and the integration in this expression
should be taken $(0,\infty)$ because for $\xi<0$ this critical point represents
a stable node). Where the result is 
$$
\Delta \ln{a} = \ln{\left(\frac{a_{1}^{\text{ini}}}{a_{s}}\right)} = -\ve(1-6\xi)
z(\lambda^{*}_{1})z'(\lambda^{*}_{1})(\lambda^{\text{ini}}_{1}-\lambda^{*}_{1})
$$
Finally we receive
$$
a_{1}^{\text{ini}} = a_{s} \exp{\big\{-\ve(1-6\xi)
z(\lambda^{*}_{1})z'(\lambda^{*}_{1})(\lambda^{\text{ini}}_{1}-\lambda^{*}_{1})\big\}}
$$
where the value in the exponent is finite, and $a_{1}^{\text{ini}}$ is the value of
the scale factor at $\tau=0$ and $a_{s}$ is the value of the scale factor at
singularity. This equation gives us the scale
factor growth from singularity to the some initial point where linear
approximation is still valid. From simple considerations we have that
$$
z(\lambda^{*}_{1})z'(\lambda^{*}_{1})\big(\lambda^{\text{ini}}_{1}-\lambda^{*}_{1}\big)<0
$$
which show that the critical point under consideration represents the finite
scale
factor singularity. Moreover it is a past singularity for the canonical scalar
field with $0<\xi<1/6$ and the phantom scalar field with $\xi>1/6$ and future
singularity for the phantom scalar field with $\xi<0$.

\begin{figure}
\begin{center}
\epsfig{file=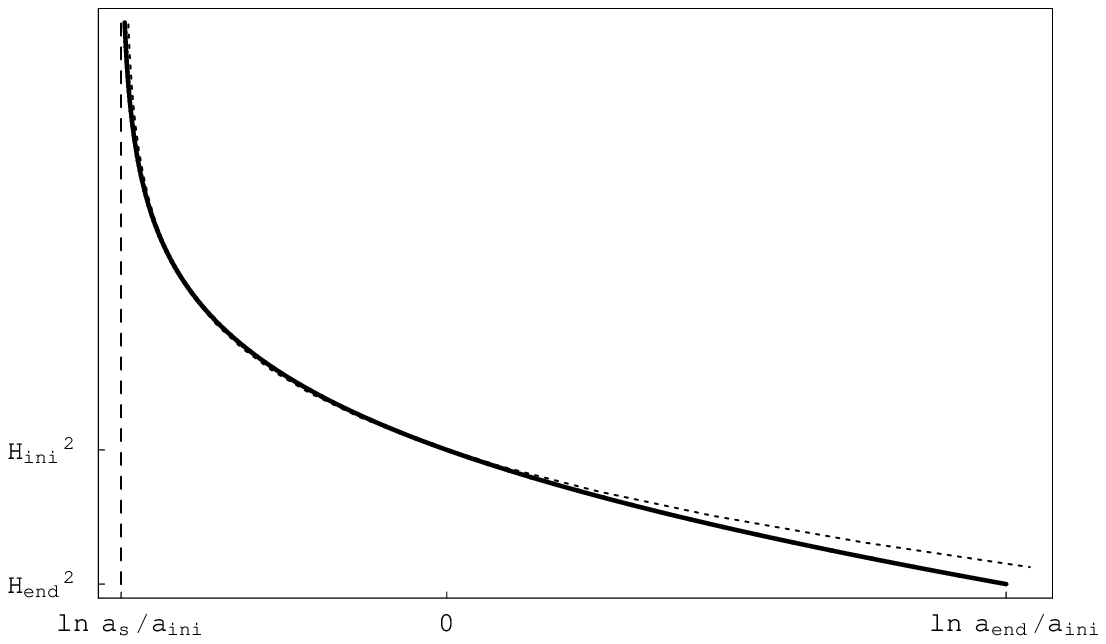,scale=0.65}
\epsfig{file=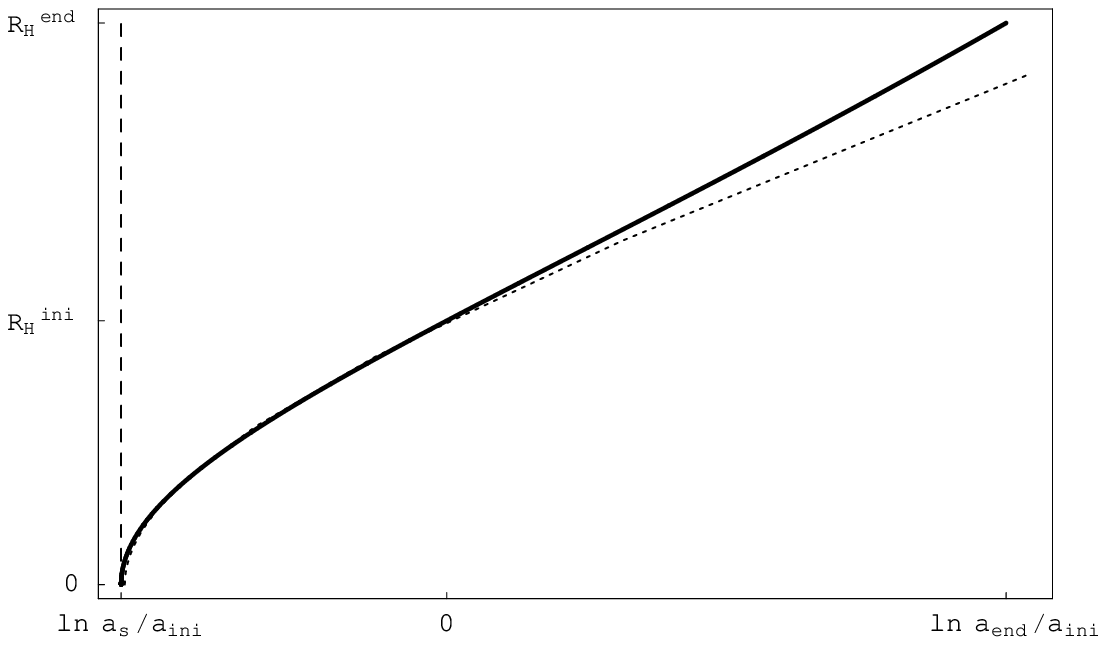,scale=0.65}
\end{center}
\caption{Evolution of $\ln{H^{2}}$ (left panel) and $R_{H}$ (right panel) as a
function of a natural logarithm of the scale factor $\ln{a}$ for a sample
trajectory with $\ve=+1$, $\xi=\frac{1}{8}$,
$z'(\lambda^{*}_{1})=\frac{1}{\alpha}=100$ in the
vicinity of the critical point corresponding to the finite scale factor
singularity. The solid black line represents the linearised solution
(\ref{eq:linsing}) and the dotted line corresponds to the numerical solution of the
system (\ref{eq:dynsys}).}
\label{fig:lin1}
\end{figure}

Now, using linearised solutions (\ref{eq:sol1}), we can express
(\ref{eq:timerep}) and (\ref{eq:H2}) as parametric functions of time $\tau$
\begin{equation}
\left\{\begin{array}{lcl}
\ln{\left(\frac{a}{a_{1}^{\text{ini}}}\right)} & = &
\ve(1-6\xi)z(\lambda_{1}^{*})z'(\lambda_{1}^{*})\big(\lambda_{1}^{\text{ini}}-\lambda_{1}^{*}\big)
\big(1-\exp{(l_{3}\tau)}\big) ,\\
\ln{\left(\frac{H}{H_{1}^{\text{ini}}}\right)^{2}} &=& 3\Big\{-4\xi\tau
-\ve\frac{4}{3}(1-6\xi)z(\lambda_{1}^{*})[(x_{1}^{\text{ini}}-x_{1}^{*})
+2(1-3\xi)z'(\lambda_{1}^{*})\big(\lambda_{1}^{\text{ini}}-\lambda_{1}^{*}\big)]\big(1-\exp{(l_{1}\tau)}\big)
\\ & &
+
\ve\frac{1}{3}(1-6\xi)(1-3w_{m}-12\xi)z(\lambda_{1}^{*})z'(\lambda_{1}^{*})\big(\lambda_{1}^{\text{ini}}-\lambda_{1}^{*}\big)\big(1-\exp{(l_{3}\tau)}\big)\Big\}.
\end{array}\right.
\label{eq:linsing}
\end{equation}
The linearised solutions used to obtain these relations are valid up to the Lyapunov
characteristic time which is equal to the inverse of the largest eigenvalue of
the linearization matrix. In our case it is
$\tau_{\text{end}}=\frac{1}{l_{3}}=\frac{1}{12\xi}$. Inserting this in to equations
(\ref{eq:linsing}) we obtain maximal values of the scale factor and the Hubble's
function respectively:
\begin{equation}
\left\{\begin{array}{lcl}
\ln{\left(\frac{a_{1}^{\text{end}}}{a_{1}^{\text{ini}}}\right)} & = &
\ve(1-6\xi)z(\lambda_{1}^{*})z'(\lambda_{1}^{*})\big(\lambda_{1}^{\text{ini}}-\lambda_{1}^{*}\big)
\big(1-e\big) ,\\
\ln{\left(\frac{H_{1}^{\text{end}}}{H_{1}^{\text{ini}}}\right)^{2}} &=& 3\Big\{-\frac{1}{3}
-\ve\frac{4}{3}(1-6\xi)z(\lambda_{1}^{*})[(x_{1}^{\text{ini}}-x_{1}^{*})
+2(1-3\xi)z'(\lambda_{1}^{*})\big(\lambda_{1}^{\text{ini}}-\lambda_{1}^{*}\big)]\big(1-e\big)
\\ & &
+
\ve\frac{1}{3}(1-6\xi)(1-3w_{m}-12\xi)z(\lambda_{1}^{*})z'(\lambda_{1}^{*})\big(\lambda_{1}^{\text{ini}}-\lambda_{1}^{*}\big)\big(1-e\big)\Big\}.
\end{array}\right.
\end{equation}
The plot representing the evolution of these quantities together with Hubble's
horizon is presented in figure~\ref{fig:lin1}. As one can simply conclude 
$$
\xi>0 \qquad : \qquad \lim_{\tau\to-\infty} H^{2} \to \infty.
$$

The general conclusion is that any phantom scalar field cosmological model with the 
negative coupling constant $\xi<0$ and the potential function which can be
represented by function $z(\lambda)$ possesses the finite scale factor future
singularity with $w_{\rm{eff}}=\pm\infty$.

\subsection{Inflation with the non-minimal coupling and arbitrary potential}

Now we proceed to the very important phase of the evolution of the universe,
namely the inflation.

\subsubsection{Fast-roll inflation}

The critical point located at
\begin{equation}
\label{eq:cp2a}
x^{*}_{2a} = -6\xi z(\lambda^{*}_{2a}), \quad (y^{*}_{2a})^{2} =
\frac{4\xi}{2\xi\lambda^{*}_{2a}z(\lambda^{*}_{2a})+(1+w_{m})}, \quad
\lambda^{*}_{2a} :
z(\lambda)^{2} = \frac{1}{\ve6\xi(1-6\xi)}
\end{equation}
with
$$
w_{\textrm{eff}} = w_{m} - 4\xi
$$
we identify as a fast-roll inflation (or rapid-roll) \cite{Linde:2001ae,
Kofman:2007tr, Chiba:2008ia}. The first reason is that $w_{\rm{eff}}$ calculated at this critical
point can be made close to $-1$ especially for the phantom scalar field, and the second
one is that the first coordinate of this point, using transformations
(\ref{eq:envar}) can be put in the following form
$$
\dot{\phi}=-6\xi H \phi
$$
which for the conformal coupling $\xi=1/6$, reduces to condition for the
rapid-roll inflation given by Kofman and Mukohyama in \cite{Kofman:2007tr}.
That is we identify this critical point as a generalisation to the non-minimally
coupled case (both for the canonical and phantom scalar fields) with additional
presence of the barotropic matter with the equation of state parameter $w_{m}$.

The linearization matrix for this critical point is the following
\begin{equation}
\label{eq:lin2a}
A_{2a}= \left(
\begin{array}{ccc}
0  & 0 & \frac{\partial x'}{\partial \lambda}\big|_{2a} \\
\frac{\partial y'}{\partial x}\big|_{2a} & -12\xi& \frac{\partial y'}{\partial
\lambda}\big|_{2a} \\
0 & 0 & 12\xi \\
\end{array} \right),
\end{equation}
where
$$
\begin{array}{ccl}
\frac{\partial x'}{\partial \lambda}\big|_{2a} & = & -3 (y^{*}_{2a})^{2}
\Big(1+w_{m}+4\xi(1-3\xi)\lambda^{*}_{2a}z(\lambda^{*}_{2a})\Big)z'(\lambda^{*}_{2a}),
\\
\frac{\partial y'}{\partial x}\big|_{2a} & = &
-\ve12\xi(1-6\xi)y^{*}_{2a}z(\lambda^{*}_{2a}),\\
\frac{\partial y'}{\partial\lambda}\big|_{2a} & = & -3\xi y^{*}_{2a}\Big[
(y^{*}_{2a})^{2}z(\lambda^{*}_{2a})+\Big(\ve6(1-6\xi)(1+w_{m})z(\lambda^{*}_{2a})+(2+(y^{*}_{2a})^{2})\lambda^{*}_{2a}\Big)z'(\lambda^{*}_{2a})\Big]
\end{array}
$$
The eigenvalues of the linearization matrix are obviously $l_{1}=0$,
$l_{2}=12\xi$ and $l_{3}=-12\xi$. Thus the fixed point is a non-hyperbolic and we
cannot make any conclusions concerning its stability based on linearization and
the Hartman-Grobman theorem is not applicable \cite{Perko:2001,Wiggins:2003}.
The answer to the question of stability or instability lies in the center
manifold theory (see appendix \ref{appa}).

We apply following procedure: first, we expand the right hand side of the
dynamical system (\ref{eq:dynsys}) into the Taylor series around the critical
point (\ref{eq:cp2a}) up to second order, and second, we make following change
of dynamical variables
$$
\left(\begin{array}{c} u \\ v \\ w \end{array}\right)=P^{-1}_{2a}\left(\begin{array}{c} x-x^{*}_{2a} \\ y-y^{*}_{2a} \\
\lambda-\lambda^{*}_{2a}\end{array}\right),
$$
where the matrix $P_{2a}$ is constructed from eigenvectors of the linearization matrix
(\ref{eq:lin2a}) calculated for corresponding eigenvalues and its inverse is
$$
P^{-1}_{2a} = \left(\begin{array}{ccc}
1 & 0 & -\frac{1}{12\xi}\frac{\partial x'}{\partial\lambda}\big|_{2a}\\
-\frac{1}{12\xi}\frac{\partial y'}{\partial x}\big|_{2a} & 1 &
-\frac{1}{24\xi}\left(-\frac{1}{12\xi}\frac{\partial
x'}{\partial\lambda}\big|_{2a} \frac{\partial y'}{\partial x}\big|_{2a} +
\frac{\partial y'}{\partial\lambda}\big|_{2a}\right) \\
0 & 0 & 1 \end{array}\right).
$$
Then dynamical system in the vicinity of the critical point representing the
fast-roll inflation is in the following form
\begin{subequations}
\begin{eqnarray}
u' & = &- 3y^{*}_{2a}\Big(2\xi\lambda^{*}_{2a}z(\lambda^{*}_{2a})+1+w_{m}\Big)uv
+ A_{u}w^{2} + B_{u} v w + C_{u} u w,\\
v' & = & -12\xi v +\ve\frac{1}{2}(1-3w_{m}) y^{*}_{2a} u^{2}
-\frac{9}{2}\Big(2\xi\lambda^{*}_{2a}z(\lambda^{*}_{2a})+1+w_{m}\Big)y^{*}_{2a}
v^{2} +\ve12\xi(1-6\xi)z(\lambda^{*}_{2a})uv \nonumber \\ & & + A_{v}w^{2} + B_{v}vw +C_{v}uw,\\
w' & = & 12\xi w + A_{w} w^{2} + B_{w} u w,
\end{eqnarray}
\end{subequations}
where $A_{i}$, $B_{i}$ and $C_{i}$ are coefficients consisting of second
derivatives of right-hand sides of dynamical system (\ref{eq:dynsys}) calculated
at the critical point under considerations.
\begin{figure}
\begin{center}
\epsfig{file=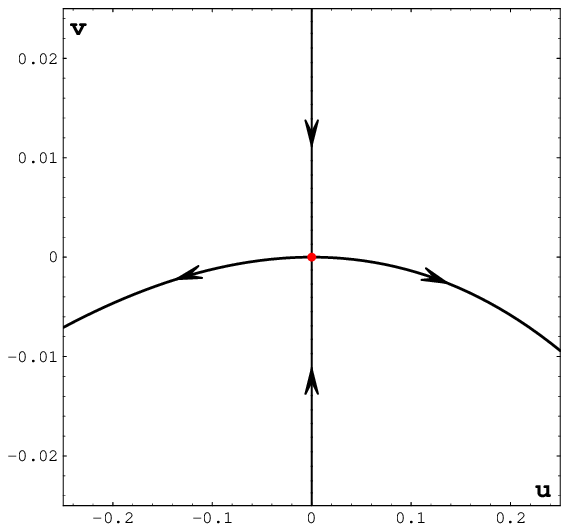,scale=1}
\epsfig{file=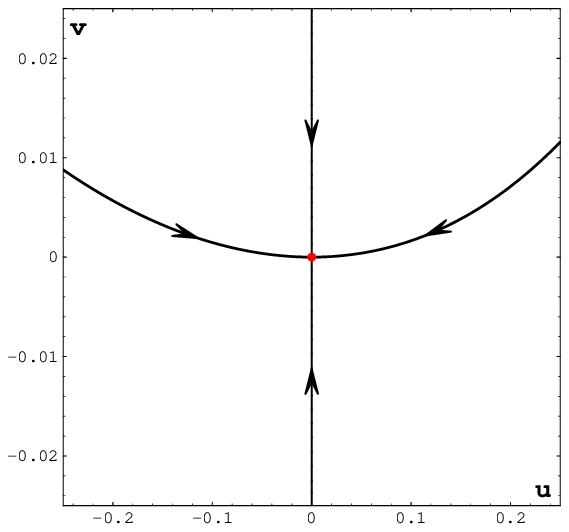,scale=1}
\end{center}
\caption{The phase portrait for the system (\ref{eq:dyn_cp2a}) on the invariant
submanifold
in the vicinity of the critical point corresponding to the rapid-roll
inflation. The bold line represents the center manifold for the problem. On the
left diagram we present an unstable case for $\ve(1-3w_{m})<0$ and the right
diagram for a stable case for $\ve(1-3w_{m})>0$. The example is given for
$z(\lambda)=\frac{\lambda}{\alpha}$ and $\alpha=1$, $w_{m}=0$ left for $\ve=-1$
and $\xi=1/4$, right for $\ve=1$ and $\xi=1/8$.}
\label{fig:1}
\end{figure}
We can note that this dynamical system admits the invariant submanifold $w=0$,
and the
dynamics can be well approximated on this submanifold. Then
\begin{subequations}
\label{eq:dyn_cp2a}
\begin{eqnarray}
\label{eq:dyn_cp2a:a}
u'&=&-3y^{*}_{2a}\Big(2\xi\lambda^{*}_{2a}z(\lambda^{*}_{2a})+1+w_{m}\Big)uv, \\
v' & = & -12\xi v +\ve\frac{1}{2}(1-3w_{m}) y^{*}_{2a} u^{2}
-\frac{9}{2}\Big(2\xi\lambda^{*}_{2a}z(\lambda^{*}_{2a})+1+w_{m}\Big)y^{*}_{2a}
v^{2} +\ve12\xi(1-6\xi)z(\lambda^{*}_{2a})uv,
\end{eqnarray}
\end{subequations}
on the invariant submanifold $w=0$.

From the center manifold theorem (appendix \ref{appa}) we have
$$
v=h(u) = \ve\frac{1}{24\xi}(1-3w_{m})y^{*}_{2a} u^{2} +
\frac{1-6\xi}{24\xi}(1-3w_{m})z(\lambda^{*}_{2a})y^{*}_{2a} u^{3} + O(u^{4})
$$
and inserting this approximation into (\ref{eq:dyn_cp2a:a}) we receive that the
vector field restricted to the center manifold is given by
$$
\eta' = -\ve\frac{1}{2}(1-3w_{m})\eta^{3} + O(\eta^{4}),
$$
which indicates that for $\ve(1-3w_{m})<0$ it is an unstable and for
$\ve(1-3w_{m})>0$ it is a stable critical point on the invariant submanifold
$w=0$
(see figure \ref{fig:1} example for $z(\lambda)=\frac{\lambda}{\alpha}$). This
equation can be simply integrated resulting in
$$
\eta(\tau)^{2} = \frac{(\eta^{\text{ini}})^{2}}{(\eta^{\text{ini}})^{2}\ve(1-3w_{m})\tau+1}.
$$
Above equation describes behaviour of the system on the center manifold which
constitutes the invariant submanifold. This solution can be used in
construction of exact solution of the system (\ref{eq:dyn_cp2a}) in the vicinity
of the critical point representing the fast-roll inflation epoch.

\begin{figure}
\begin{center}
\epsfig{file=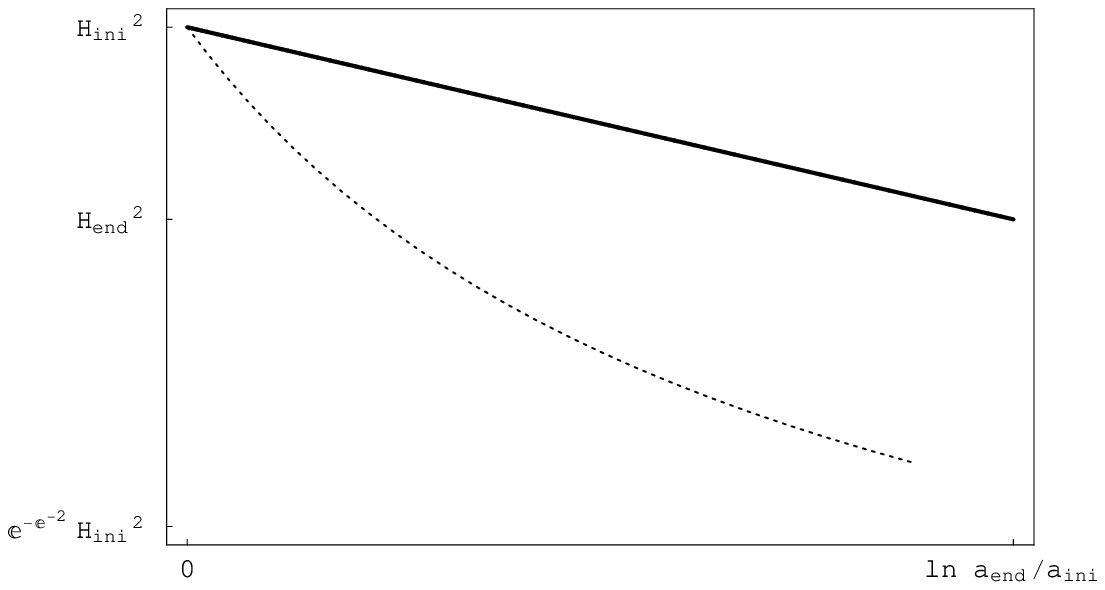,scale=0.64}
\epsfig{file=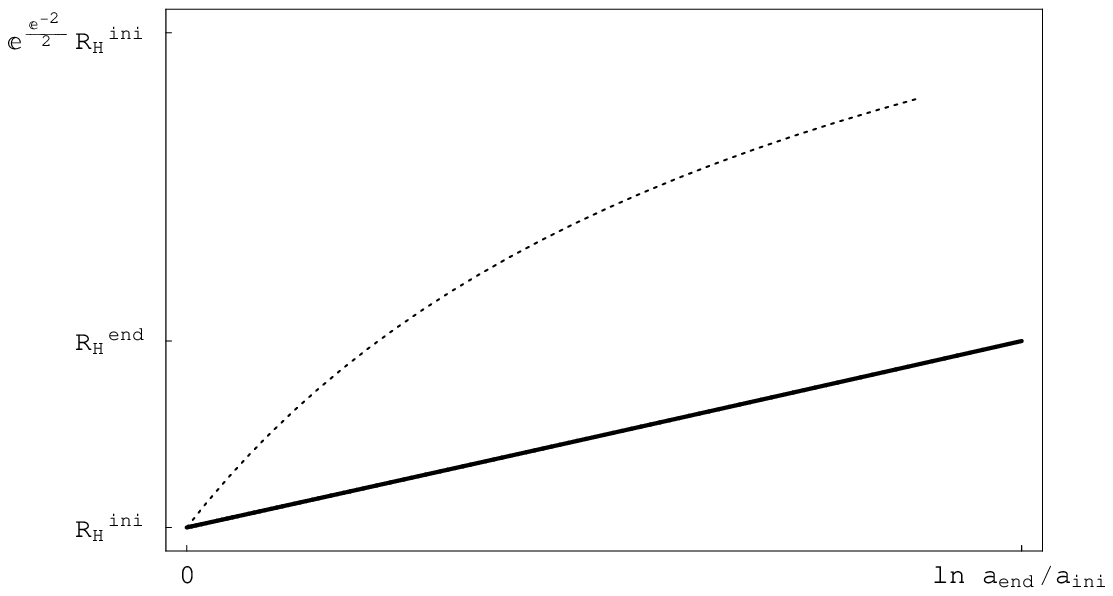,scale=0.64}
\end{center}
\caption{Evolution of $\ln{H^{2}}$ (left panel) and $R_{H}$ (right panel) as a
function of natural logarithm of the scale factor $\ln{a}$ for a sample
trajectory with $\ve=-1$, $\xi=\frac{1}{4}$,
$z'(\lambda^{*}_{1})=\frac{1}{\alpha}=10$ in the
vicinity of the critical point corresponding to the fast-roll inflation. The
solid black line represents the linearised solution (\ref{eq:H2lin2a}) and the
dotted line represents the numerical solution of the system (\ref{eq:dynsys}).}
\label{fig:lin2a}
\end{figure}

Using the solution from the center manifold theorem and keeping linear term in
$w$ only
$$
\begin{array}{c}
u(\tau)v(\tau)\propto u(\tau)^{3}\approx0, \quad v(\tau)w(\tau)\propto
u(\tau)^{2}w(\tau)\approx0, \quad v(\tau)^{2}\propto u(\tau)^{4}\approx0, \quad
w(\tau)^{2}\approx0, \quad u(\tau)w(\tau)\approx0
\end{array}
$$
from (\ref{eq:timerep}) and (\ref{eq:H2}) we get the parametric equations for
the evolution of the scale factor and Hubble's function
\begin{equation}
\left\{\begin{array}{lcl}
\ln{\left(\frac{a}{a_{2a}^{\text{ini}}}\right)} & = &
-\ve(1-6\xi)z(\lambda_{2a}^{*})z'(\lambda_{2a}^{*})w^{\text{ini}}\Big(\exp{(12\xi\tau)}-1\Big),\\
\ln{\left(\frac{H}{H_{2a}^{\text{ini}}}\right)^{2}} & = & -\frac{1}{4\xi} A
w^{\text{ini}}\Big(\exp{(12\xi\tau)}-1\Big),
\end{array}\right.
\label{eq:H2lin2a}
\end{equation}
which can be easy combine resulting in
\begin{equation}
\ln{\left(\frac{H}{H_{2a}^{\text{ini}}}\right)^{2}} =
-\frac{A}{\ve4\xi(1-6\xi)z(\lambda_{2a}^{*})z'(\lambda_{2a}^{*})}
\ln{\left(\frac{a}{a_{2a}^{\text{ini}}}\right)}
\end{equation}
where
$$
\begin{array}{ll}
A = \frac{1}{72\big(2\xi\lambda_{2a}^{*}z(\lambda_{2a}^{*})+(1+w_{m})\big)}\Big\{
\big[& -\ve144\xi(1-6\xi)(1+w_{m})(1+3w_{m})z(\lambda_{2a}^{*}) +
96\xi(2-9\xi)\lambda_{2a}^{*}+\\
&+288\xi^{2}(\lambda_{2a}^{*})^{2}z(\lambda_{2a}^{*})\big]z'(\lambda_{2a}^{*})-288\xi^{2}z(\lambda_{2a}^{*})\Big\}.
\end{array}
$$
One can conclude that needs $|A|\ll1$ in order to achieve $H^{2}\approx \text{const.}$ during the
evolution. The linearised solution in $w$ direction is valid up to the Lyapunov
time $\tau_{end}=\frac{1}{12\xi}$, using this we obtain maximal values of the
scale factor and the Hubble's function respectively:
\begin{equation}
\left\{\begin{array}{lcl}
\ln{\left(\frac{a_{2a}^{\text{end}}}{a_{2a}^{\text{ini}}}\right)} & = &
-\ve(1-6\xi)z(\lambda_{2a}^{*})z'(\lambda_{2a}^{*})w^{\text{ini}}\Big(e-1\Big),\\
\ln{\left(\frac{H_{2a}^{\text{end}}}{H_{2a}^{\text{ini}}}\right)^{2}} & = & -\frac{1}{4\xi} A
w^{\text{ini}}\Big(e-1\Big),
\end{array}\right.
\end{equation}
In figure \ref{fig:lin2a} we present the evolution of Hubble's function and
Hubble's horizon in the vicinity of this critical point.

\subsubsection{Slow-roll inflation}

The critical point located at
\begin{equation}
\label{eq:cp2b}
x^{*}_{2b} = 0, \quad (y^{*}_{2b})^{2} =
\frac{2\xi(1-3w_{m})}{(1-6\xi)\big(2\xi\lambda^{*}_{2b}z(\lambda^{*}_{2b})+(1+w_{m})\big)},
\quad \lambda^{*}_{2b} \colon z(\lambda)^{2} = \frac{1}{\ve6\xi(1-6\xi)}
\end{equation}
with
$$
w_{\text{eff}} = \frac{w_{m}-2\xi}{1-6\xi}
$$
we identify as representing the phase of a slow-roll inflation due to
$x\propto \dot{\phi}$ so the dynamics in the vicinity of
this point corresponds to the slow-roll condition $\dot{\phi}\approx0$.

The linearization matrix is in the form
\begin{equation}
A_{2b}= \left(
\begin{array}{ccc}
\frac{\partial x'}{\partial x}\big|_{2b}  & \frac{\partial x'}{\partial y}\big|_{2b} &
\frac{\partial x'}{\partial \lambda}\big|_{2b} \\
 \frac{\partial y'}{\partial x}\big|_{2b} & \frac{\partial y'}{\partial
 y}\big|_{2b} &
 \frac{\partial y'}{\partial \lambda}\big|_{2b} \\
 0 & 0 & 0 \\
\end{array} \right),
\end{equation}
where nonzero elements are
$$
\begin{array}{ccl}
\frac{\partial x'}{\partial x}\big|_{2b} & = & \frac{6\xi}{1-6\xi}(1-3w_{m}),\\
\frac{\partial x'}{\partial y}\big|_{2b} & = &
-18\xi\Big(2\xi\lambda^{*}_{2b}z(\lambda^{*}_{2b})+
(1+w_{m})\Big)y^{*}_{2b}z(\lambda^{*}_{2b}), \\
\frac{\partial x'}{\partial \lambda}\big|_{2b} & = &
\frac{3}{1-6\xi}(y^{*}_{2b})^{2}
\Big(-\ve\xi+(1-6\xi)z'(\lambda^{*}_{2b})\big(6\xi^{2}\lambda^{*}_{2b}z(\lambda^{*}_{2b})+(1+w_{m})\big)
\Big), \\
\frac{\partial y'}{\partial x}\big|_{2b} & = &
\ve6\xi(1-3w_{m})y^{*}_{2b}z(\lambda^{*}_{2b}),\\
\frac{\partial y'}{\partial y}\big|_{2b} & = & -\frac{6\xi}{1-6\xi}(1-3w_{m}),\\
\frac{\partial y'}{\partial \lambda}\big|_{2b} & = & -3\xi
y^{*}_{2b}\Big((y^{*}_{2b})^{2} z(\lambda^{*}_{2b}) +
z'(\lambda^{*}_{2b})\big((y^{*}_{2b})^{2}\lambda^{*}_{2b}+\ve 6
z(\lambda^{*}_{2b})(1+w_{m}-8\xi)\big)\Big),
\end{array}
$$
and additionally we have the following relation
$$\frac{\partial x'}{\partial x}\Big|_{2b}\ \frac{\partial y'}{\partial
y}\Big|_{2b}  =  - \left(\frac{6\xi}{1-6\xi}\right)^{2}(1-3w_{m})^{2}.$$
The characteristic equation for the linearization matrix gives us vanishing
eigenvalues $l_{1}=l_{2}=l_{3}=0$, so the critical point is degenerated. In this
case we cannot use standard procedures, following the Hartman-Grobman theorem
\cite{Perko:2001,Wiggins:2003} of determining qualitative behaviour of
the investigated system in the vicinity of this critical point.
Instead we can notice that the linearization matrix $A_{2b}$ calculated at this
critical point is nilpotent of order $3$, i.e. $(A_{2b})^{3}=0$. Then solution
of the linearised problem can be presented in the following form
$$
\mathbf{x}(\tau) = \Big[1+A_{2b}\tau + \frac{1}{2}(A_{2b})^{2}\tau^{2}\Big] \mathbf{x}_{0}
$$

Finally solutions in the vicinity of this degenerated critical point up to
linear terms are
\begin{equation}
\begin{array}{lcl}
x(\tau) & = & x^{\text{ini}}_{2b} + \bigg(\frac{\partial x'}{\partial
x}\Big|_{2b}(x^{\text{ini}}_{2b}-x^{*}_{2b}) + \frac{\partial x'}{\partial
y}\Big|_{2b}(y^{\text{ini}}_{2b}-y^{*}_{2b}) + \frac{\partial x'}{\partial
\lambda}\Big|_{2b}(\lambda^{\text{ini}}_{2b}-\lambda^{*}_{2b})\bigg)\tau \\
& & + \frac{1}{2} \bigg(\frac{\partial x'}{\partial x}\Big|_{2b}\frac{\partial
x'}{\partial \lambda}\Big|_{2b} + \frac{\partial x'}{\partial
y}\Big|_{2b}\frac{\partial y'}{\partial \lambda}\Big|_{2b}\bigg)
(\lambda^{\text{ini}}_{2b}-\lambda^{*}_{2b})\tau^{2}, \\
y(\tau) & = & y^{\text{ini}}_{2b} + \bigg(\frac{\partial y'}{\partial
x}\Big|_{2b}(x^{\text{ini}}_{2b}-x^{*}_{2b}) + \frac{\partial y'}{\partial
y}\Big|_{2b}(y^{\text{ini}}_{2b}-y^{*}_{2b}) + \frac{\partial y'}{\partial
\lambda}\Big|_{2b}(\lambda^{\text{ini}}_{2b}-\lambda^{*}_{2b})\bigg)\tau \\
& & + \frac{1}{2}\bigg(\frac{\partial y'}{\partial y}\Big|_{2b}\frac{\partial
y'}{\partial \lambda}\Big|_{2b} + \frac{\partial y'}{\partial
x}\Big|_{2b}\frac{\partial x'}{\partial \lambda}\Big|_{2b}\bigg)
(\lambda^{\text{ini}}_{2b}-\lambda^{*}_{2b})\tau^{2}, \\
\lambda(\tau) & = & \lambda^{\text{ini}}_{2b}.
\end{array}
\label{eq:lin2b}
\end{equation}
where
$$
\begin{array}{ccl}
\frac{\partial x'}{\partial x}\big|_{2b}\frac{\partial
x'}{\partial \lambda}\big|_{2b} + \frac{\partial x'}{\partial
y}\big|_{2b}\frac{\partial y'}{\partial \lambda}\big|_{2b} & = &
36\xi(1+w_{m})\big(2+3\xi\lambda^{*}_{2b} z(\lambda^{*}_{2b})\big)
(y^{*}_{2b})^{2} z'(\lambda^{*}_{2b}),\\
\frac{\partial y'}{\partial y}\big|_{2b}\frac{\partial
y'}{\partial \lambda}\big|_{2b} + \frac{\partial y'}{\partial
x}\big|_{2b}\frac{\partial x'}{\partial \lambda}\big|_{2b} & = & 18\xi
(1+w_{m})\big(\lambda^{*}_{2b}+\ve4(1-6\xi)z(\lambda^{*}_{2b})\big)
(y^{*}_{2b})^{3}z'(\lambda^{*}_{2b}).
\end{array}
$$
These linearised solutions are valid up to a maximal value of the time parameter
$
\tau = \tau_{\rm{max}}
$
which can be used to calculate the scale factor growth during the slow roll
inflation
$$
\ln{\frac{a^{\text{end}}_{si}}{a^{\text{start}}_{si}}} =
-\ve24\xi(1-6\xi)z(\lambda^{*}_{2b})z'(\lambda^{*}_{2b})
\big(\lambda^{\text{ini}}_{2b}-\lambda^{*}_{2b}\big)\tau_{\rm{max}}.
$$

\begin{figure}
\begin{center}
\epsfig{file=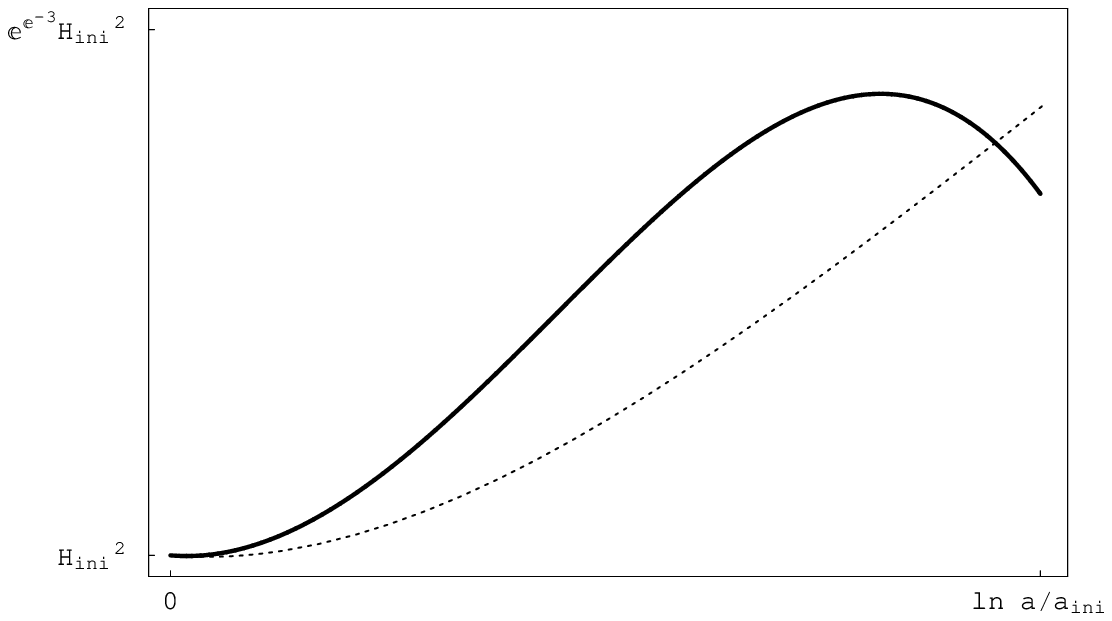,scale=0.64}
\epsfig{file=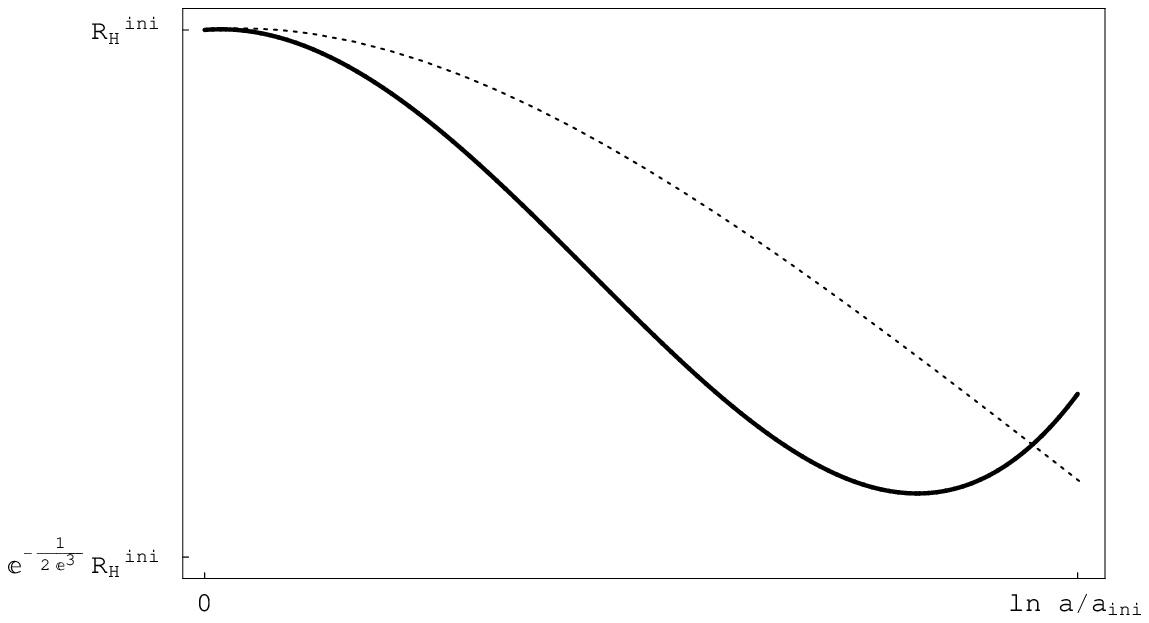,scale=0.64}
\end{center}
\caption{Evolution of $\ln{H^{2}}$ (left panel) and $R_{H}$ (right panel) as a
function of the natural logarithm of the scale factor $\ln{a}$ for a sample
trajectory with $\ve=+1$, $\xi=\frac{1}{8}$,
$z'(\lambda^{*}_{1})=\frac{1}{\alpha}=100$ in the
vicinity of the critical point corresponding to the slow-roll inflation. The
solid black line represents the linearised solution (\ref{eq:lin2b}) and the
dotted line represents the numerical solution of the system (\ref{eq:dynsys}).}
\label{fig:lin2b}
\end{figure}

The direct application of the linearised solutions (\ref{eq:lin2b}) to
(\ref{eq:timerep}) and (\ref{eq:H2}) gives us the approximated evolution of
Hubble's function in the vicinity of the critical point representing the
slow-roll inflation (figure \ref{fig:lin2b}).

\subsection{Radiation domination epoch generated by non-minimal coupling}

Following critical point located at
\begin{equation}
\label{eq:cp3a}
x^{*}_{3a} \colon g(x)=0, \quad y^{*}_{3a}=0, \quad \lambda^{*}_{3a} \colon z(\lambda)^{2} =
\frac{1}{\ve6\xi(1-6\xi)}
\end{equation}
where $g(x)=\ve(1-4\xi-w_{m})x^{2}+\ve4\xi(1-3w_{m})z(\lambda^{*}_{3a})x
+\frac{2\xi}{1-6\xi}(1-3w_{m})$ and at this point value of the effective equation of
the state parameter is
$$
w_{\text{eff}} = \frac{1}{3}
$$
represents the radiation dominated universe. With solutions to $g(x)=0$
equation in th form
$$
x_{1,2}=\frac{1}{\ve2(1-4\xi-w_{m})}\left\{-\ve4\xi(1-3w_{m})z(\lambda^{*}_{3a})
\pm \sqrt{-\ve\frac{16}{3}\xi(1-3w_{m})}\right\}
$$
which is real only if the expression in square root is positive and for the
barotropic matter with $w_{m}<\frac{1}{3}$ it is possible only if $\ve\xi<0$. We
are interested only in evolution with $\xi>0$ because of the discussion of the
critical point representing the finite scale factor singularity, and this is the
reason we identify this critical point as a representing radiation dominated
epoch only for the phantom scalar field.

The linearization matrix calculated at this point is in the following form
\begin{equation}
\label{eq:lin3a}
A_{3a}= \left(
\begin{array}{ccc}
\frac{\partial x'}{\partial x}\big|_{3a} & 0 & \frac{\partial x'}{\partial
\lambda}\big|_{3a} \\
0 & 0 & 0 \\
0 & 0 & \frac{\partial \lambda'}{\partial \lambda}\big|_{3a} \\
\end{array} \right),
\end{equation}
where
$$
\begin{array}{ccl}
\frac{\partial x'}{\partial x}\big|_{3a} & = &
\ve12\xi(1-6\xi)z(\lambda^{*}_{3a})x^{*}_{3a}, \\
\frac{\partial x'}{\partial \lambda}\big|_{3a} & = & \ve 6\xi
z'(\lambda^{*}_{3a}) \Big[2(1-6\xi)z(\lambda^{*}_{3a})x^{*}_{3a}
+(1-3w_{m})\big(x^{*}_{3a}+6\xi z(\lambda^{*}_{3a})\big)
\big(x^{*}_{3a}+z(\lambda^{*}_{3a})\big)\Big], \\
\frac{\partial \lambda'}{\partial \lambda}\big|_{3a} & = &
-\ve12\xi(1-6\xi)z(\lambda^{*}_{3a})x^{*}_{3a}.
\end{array}
$$

Eigenvalues of the linearization matrix are
$l_{1}=-\ve12\xi(1-6\xi)z(\lambda^{*}_{3a})x^{*}_{3a}$, $l_{2}=0$,
$l_{3}=\ve12\xi(1-6\xi)z(\lambda^{*}_{3a})x^{*}_{3a}$. This indicates that the
critical point is non-hyperbolic one and the standard linearization procedure
will be inefficient and we need to proceed with the center manifold theorem (see
appendix \ref{appa}) and the procedure described during the discussion of the
critical
point representing fast-roll inflation. We make following change of
dynamical variables
$$
\left(\begin{array}{c} u \\ v \\ w
\end{array}\right)=P^{-1}_{3a}\left(\begin{array}{c} x-x^{*}_{3a} \\
y-y^{*}_{3a} \\
\lambda-\lambda^{*}_{3a}\end{array}\right),
$$
where matrix $P_{3a}$ is constructed from eigenvectors of the linearization
matrix (\ref{eq:lin3a}) and its inverse is
$$
P_{3a}^{-1} = \left(
\begin{array}{ccc}
0 & 0 & 1 \\
0 & 1 & 0 \\
1 & 0 & \chi\\
\end{array} \right), \qquad \rm{and} \qquad \chi= \frac{\frac{\partial x'}{\partial
\lambda}\big|_{3a}}{\frac{\partial
x'}{\partial x}\big|_{3a} - \frac{\partial\lambda'}{\partial\lambda}\big|_{3a}}.
$$
Then dynamical system can be presented in the following form
\begin{equation}
\begin{array}{ccl}
u' & = & -\ve12\xi(1-6\xi)z(\lambda^{*}_{3a})x^{*}_{3a}u + A_{u}u^{2} + B_{u}uw,
\\
v' & = & A_{v} uv + B_{v} vw,\\
w' & = & \ve12\xi(1-6\xi)z(\lambda^{*}_{3a})x^{*}_{3a}w + A_{w}u^{2} +
B_{w}v^{2} + C_{w} w^{2} + D_{w} u w,
\end{array}
\end{equation}
where $A_{i}$, $B_{i}$, $C_{i}$ and $D_{i}$ are coefficients consisting of
second derivatives of the right hand sides of dynamical system (\ref{eq:dynsys})
calculated at the critical point (\ref{eq:cp3a}).

One can note that above dynamical system admits two invariant submanifolds
namely $v=0$ and $u=0$.

On the first invariant submanifold the system can be simply reduced to
\begin{equation}
\begin{array}{ccl}
u' & = & -\ve12\xi(1-6\xi)z(\lambda^{*}_{3a})x^{*}_{3a}u\\
w' & = & \ve12\xi(1-6\xi)z(\lambda^{*}_{3a})x^{*}_{3a}w ,
\end{array}
\end{equation}
resulting in the solution representing a saddle type critical point in the form
\begin{equation}
\label{eq:cp3av}
\begin{array}{lcl}
u(\tau) & = & u^{\text{ini}}
\exp{\big(-\ve12\xi(1-6\xi)z(\lambda^{*}_{3a})x^{*}_{3a}\tau\big)}, \\
w(\tau) & = & w^{\text{ini}}
\exp{\big(\ve12\xi(1-6\xi)z(\lambda^{*}_{3a})x^{*}_{3a}\tau\big)}.
\end{array}
\end{equation}


\begin{figure}
\begin{center}
\epsfig{file=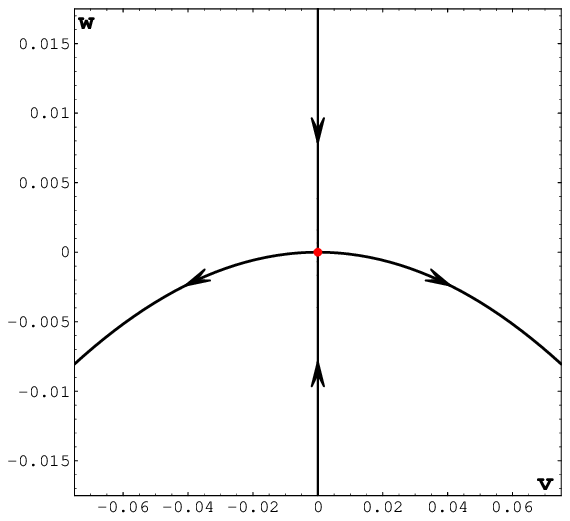,scale=1}
\epsfig{file=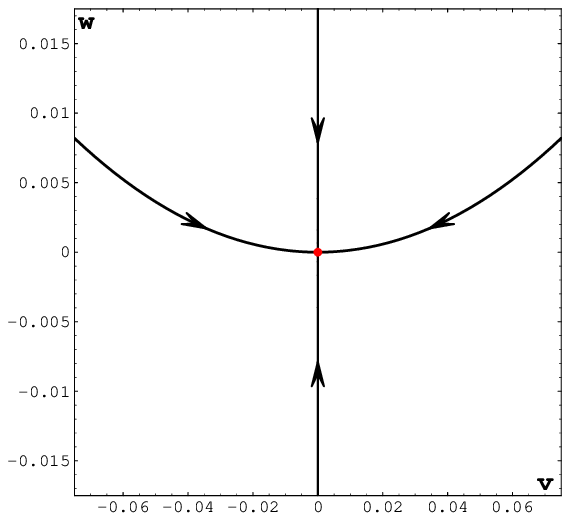,scale=1}
\end{center}
\caption{The phase portrait for the system (\ref{eq:dyn_cp3a}) on the invariant
submanifold $u=0$ in the vicinity of the critical point corresponding to the
radiation dominated universe with $w_{\rm{eff}}=\frac{1}{3}$. The bold parabola
shaped line represents the center submanifold for the problem. On the left
diagram
we present an unstable case and on the right diagram for a stable case.
The example is given for $z(\lambda)=\frac{\lambda}{\alpha}$, $\ve=-1$,
$w_{m}=0$, $\xi=1/4$ and $\alpha=1$ (left) and $\alpha=-4$ (right).}
\label{fig:2}
\end{figure}

On the other hand we can also restrict our system to the invariant submanifold
defined by $u=0$, then
\begin{equation}
\label{eq:dyn_cp3a}
\begin{array}{ccl}
v' & = & B_{v} v w ,\\
w' & = & \frac{\partial x'}{\partial x}\Big|_{3a} w + B_{w} v^{2} + C_{w} w^{2},
\end{array}
\end{equation}
and from the center manifold theorem (see appendix \ref{appa}) we have
$$
w = h(v) = -\frac{B_{w}}{\frac{\partial x'}{\partial x}|_{3a}} v^{2} +
\frac{B_{w}^{2}}{(\frac{\partial x'}{\partial x}|_{3a})^{3}}(2 B_{v}-C_{w})v^{4}
+ O(v^{5})
$$
and inserting this approximation into first equation of the system
(\ref{eq:dyn_cp3a}) we receive that the vector field restricted to the center
manifold is given by
$$
\eta' = - \frac{B_{v}B_{w}}{\frac{\partial x'}{\partial x}|_{3a}}\eta^{3} +
O(\eta^{4})
$$
where 
$$
 \frac{B_{v}B_{w}}{\frac{\partial x'}{\partial x}|_{3a}}=
 -\ve\frac{\lambda^{*}_{3a}}{2(1-6\xi)z(\lambda^{*}_{3a})} -\frac{3}{2}(1+w_{m})
$$
and this indicates that for $\frac{B_{v}B_{w}}{\frac{\partial x'}{\partial
x}|_{3a}}<0$ it is an unstable and for $\frac{B_{v}B_{w}}{\frac{\partial
x'}{\partial x}|_{3a}}>0$ it is a stable critical point on the invariant
submanifold $u=0$ (see figure \ref{fig:2} for an example for
$z(\lambda)=\frac{\lambda}{\alpha}$).


Now we are ready to present the evolution of Hubble's function in the
vicinity of this critical point. First, we use approximated solutions
(\ref{eq:cp3av}) on the invariant submanifold $v=0$. We have
\begin{equation}
\left\{\begin{array}{lcl}
\ln{\left(\frac{a}{a_{3a}^{\text{ini}}}\right)} & = &
\frac{z'(\lambda_{3a}^{*})}{x_{3a}^{*}}
u^{\text{ini}}\Big(\exp{\big(-\ve12\xi(1-6\xi)z(\lambda_{3a}^{*})x_{3a}^{*}\tau\big)}-1\Big),\\
\ln{\left(\frac{H}{H_{3a}^{\text{ini}}}\right)^{2}} & = &
\frac{1-3w_{m}}{1-6\xi}\frac{x_{3a}^{*}+6\xi
z(\lambda_{3a}^{*})}{(x_{3a}^{*})^{2}}
w^{\text{ini}}\Big(\exp{\big(\ve12\xi(1-6\xi)z(\lambda_{3a}^{*})x_{3a}^{*}\tau\big)}-1\Big)
- \\ & & -\Big(4+\frac{1-w_{m}}{4\xi}\frac{x_{3a}^{*}}{z(\lambda_{3a}^{*})}\Big)\frac{z'(\lambda_{3a}^{*})}{x_{3a}^{*}}
u^{\text{ini}}\Big(\exp{\big(-\ve12\xi(1-6\xi)z(\lambda_{3a}^{*})x_{3a}^{*}\tau\big)}-1\Big)
\end{array}\right.
\end{equation}

On the other hand, using the solution from a center manifold and keeping only linear
terms in $u$
$$
\begin{array}{c}
w(\tau)^{2}\propto v(\tau)^{4}\approx 0, \quad u(\tau)w(\tau)\propto
u(\tau)v(\tau)^{2}\approx0, \quad u(\tau)^{2}\approx0
\end{array}
$$
we receive
\begin{equation}
\left\{\begin{array}{lcl}
\ln{\left(\frac{a}{a_{3a}^{\text{ini}}}\right)} & = &
\frac{z'(\lambda_{3a}^{*})}{x_{3a}^{*}}
u^{\text{ini}}\Big(\exp{\big(-\ve12\xi(1-6\xi)z(\lambda_{3a}^{*})x_{3a}^{*}\tau\big)}-1\Big),\\
\ln{\left(\frac{H}{H_{3a}^{\text{ini}}}\right)^{2}} & = &
-\Big(4+\frac{1-w_{m}}{4\xi}\frac{x_{3a}^{*}}{z(\lambda_{3a}^{*})}\Big)\frac{z'(\lambda_{3a}^{*})}{x_{3a}^{*}}
u^{\text{ini}}\Big(\exp{\big(-\ve12\xi(1-6\xi)z(\lambda_{3a}^{*})x_{3a}^{*}\tau\big)}-1\Big)
\end{array}\right.
\label{eq:3alin}
\end{equation}
which can be easy to combine as
\begin{equation}
\ln{\left(\frac{H}{H_{3a}^{\text{ini}}}\right)^{2}} =
-\Big(4+\frac{1-w_{m}}{4\xi}\frac{x_{3a}^{*}}{z(\lambda_{3a}^{*})}\Big)
\ln{\left(\frac{a}{a_{3a}^{\text{ini}}}\right)}.
\end{equation}
One can notice that this expression resembles behaviour of the Hubble's function
during the pure radiation domination epoch, but with contribution coming from
non-minimal coupling. The linearised solutions are valid up to the Lyapunov time
$\tau_{\text{end}}=\frac{1}{-\ve12\xi(1-6\xi)z(\lambda_{3a}^{*})x_{3a}^{*}}>0$,
then inserting this in to the latter equations we receive maximal values of the
scale factor and Hubble's function valid in the center manifold
approximation
\begin{equation}
\left\{\begin{array}{lcl}
\ln{\left(\frac{a_{3a}^{\text{end}}}{a_{3a}^{\text{ini}}}\right)} & = &
\frac{z'(\lambda_{3a}^{*})}{x_{3a}^{*}}
u^{\text{ini}}\Big(e-1\Big),\\
\ln{\left(\frac{H_{3a}^{\text{end}}}{H_{3a}^{\text{ini}}}\right)^{2}} & = &
-\Big(4+\frac{1-w_{m}}{4\xi}\frac{x_{3a}^{*}}{z(\lambda_{3a}^{*})}\Big)\frac{z'(\lambda_{3a}^{*})}{x_{3a}^{*}}
u^{\text{ini}}\Big(e-1\Big)
\end{array}\right.
\end{equation}

\begin{figure}
\begin{center}
\epsfig{file=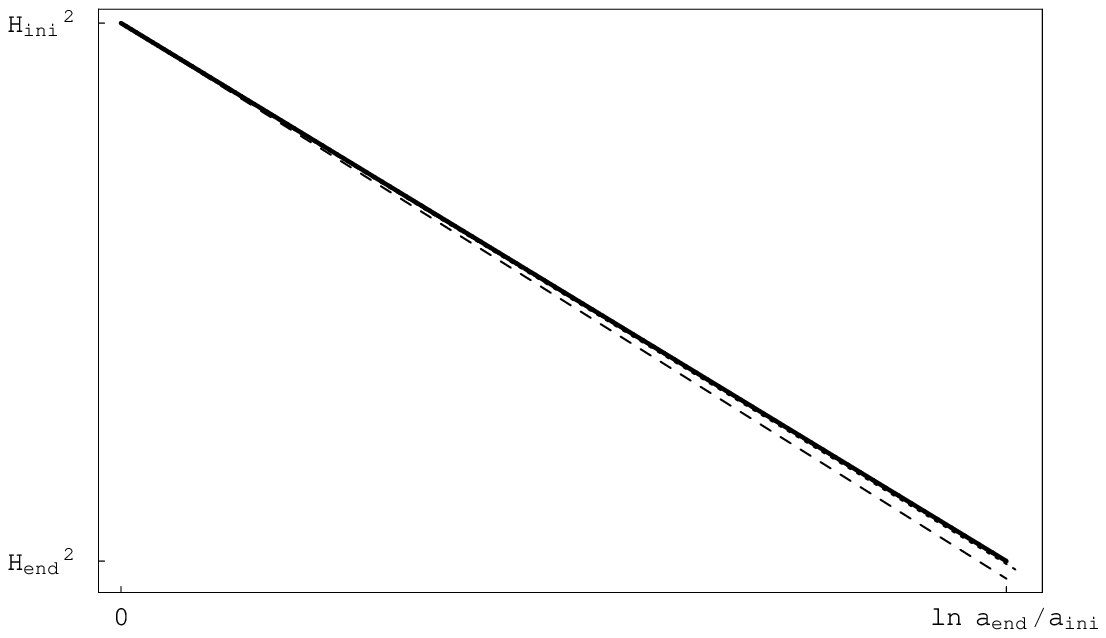,scale=0.65}
\epsfig{file=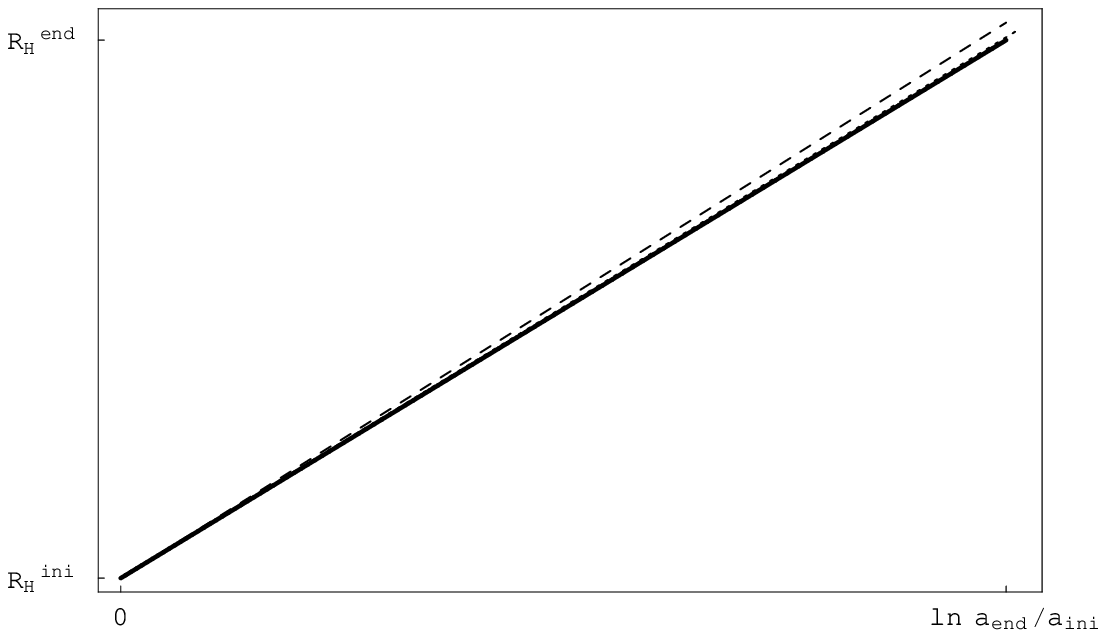,scale=0.65}
\end{center}
\caption{Evolution of $\ln{H^{2}}$ (left panel) and $R_{H}$ (right panel) for a
sample trajectory with $\ve=-1$, $\xi=10$,
$z'(\lambda^{*}_{3a})=\frac{1}{\alpha}=\frac{1}{10}$
in the vicinity of the critical point corresponding to the radiation dominated
epoch for the phantom scalar field. The solid black line corresponds to the
linearised solution (\ref{eq:3alin}), the dashed line corresponds to exact
radiation dominated expansion of the universe $\ln{H^{2}}\propto-4\ln{a}$ and
the dotted line corresponds to the numerical solution of the system
(\ref{eq:dynsys}).}
\label{fig:lin3a}
\end{figure}

In figure \ref{fig:lin3a} we present evolution of
$\ln{H^{2}}$ and $R_{H}$ as a functions of $\ln{a}$ in the vicinity of the
critical point representing radiation domination epoch for the phantom scalar field.

There is another critical point which represents the radiation dominated
universe located at
\begin{equation}
\label{eq:cp3b}
x^{*}_{3b}=0, \quad y^{*}_{3b}=0, \quad \lambda^{*}_{3b} \colon z(\lambda)^{2} =
\frac{1}{\ve6\xi}
\end{equation}
with the effective equation of the state parameter
$$
w_{\text{eff}} = \frac{1}{3}
$$
We need to stress that this critical point exists only if $w_{m}\ne\frac{1}{3}$.
Linearised solutions in the vicinity of this critical point are
\begin{subequations}
\label{eq:lin3b}
\begin{eqnarray}
x_{3b}(\tau) & = &
\frac{1-3w_{m}}{2-3w_{m}} \Big(x_{3b}^{\text{ini}} +
z'(\lambda^{*}_{3b})\big(\lambda_{3b}^{\text{ini}}-\lambda_{3b}^{*}\big)
\Big)\exp{(l_{1}\tau)} + \nonumber \\ & & +
\frac{1}{2-3w_{m}}\Big(x_{3b}^{\text{ini}}-(1-3w_{m})z'(\lambda^{*}_{3b})\big(\lambda_{3b}^{\text{ini}}-\lambda_{3b}^{*}\big)\Big)\exp{(l_{3}\tau)}, \\
y_{3b}(\tau) & = & y_{3b}^{\text{ini}}\exp{(l_{2}\tau)}, \\
\lambda_{3b}(\tau) & = & \lambda_{3b}^{*} + \frac{1}{2-3w_{m}}
\frac{1}{z'(\lambda^{*}_{3b})}
\Big(x_{3b}^{\text{ini}} +
z'(\lambda^{*}_{3b})\big(\lambda_{3b}^{\text{ini}}-\lambda_{3b}^{*}\big)
\Big)\exp{(l_{1}\tau)} - \nonumber \\ & & -
\frac{1}{2-3w_{m}} \frac{1}{z'(\lambda^{*}_{3b})}
\Big(x_{3b}^{\text{ini}}-(1-3w_{m})z'(\lambda^{*}_{3b})\big(\lambda_{3b}^{\text{ini}}-\lambda_{3b}^{*}\big)\Big)\exp{(l_{3}\tau)}.
\end{eqnarray}
\end{subequations}
where
$$
l_{1} = 6\xi(1-3w_{m}), \quad l_{2} = 12\xi, \quad l_{3}=-6\xi
$$
are the eigenvalues of the linearization matrix calculated at this critical
point. Simple inspection of this eigenvalues gives us further constraint on the
value of the barotropic matter equation of state parameter $w_{m}$, namely
$l_{1}$ should be positive resulting in $w_{m}<\frac{1}{3}$ to assure that in
($x$,$\lambda$) plane the dynamics
in the vicinity of this critical point would correspond to a saddle type
critical point. This will guarantee that the evolution proceeds towards the next critical point
representing matter dominated universe.

Using linearised solutions (\ref{eq:lin3b}) we are able to express
(\ref{eq:timerep}) and (\ref{eq:H2}) as a parametric functions of time $\tau$
\begin{equation}
\label{eq:H2lin3b}
\left\{\begin{array}{lcl}
\ln{\left(\frac{a}{a_{3b}^{\text{ini}}}\right)} & = & 6\xi\tau -
\ve2\frac{1-6\xi}{(1-3w_{m})(2-3w_{m})}z(\lambda_{3b}^{*})
\Big(x_{3b}^{\text{ini}}+z'(\lambda_{3b}^{*})\big(\lambda_{3b}^{\text{ini}}-\lambda_{3b}^{*}\big)\Big)\big(\exp{(l_{1}\tau)}-1\big)
- \\ & &
-\ve2\frac{1-6\xi}{2-3w_{m}}z(\lambda_{3b}^{*})\Big(x_{3b}^{\text{ini}}-(1-3w_{m})z'(\lambda_{3b})\big(\lambda_{3b}^{\text{ini}}-\lambda_{3b}^{*}\big)\Big)\big(\exp{(l_{3}\tau)}-1\big),\\

\ln{\left(\frac{H}{H^{\text{ini}}_{3b}}\right)^{2}} & = & -24\xi\tau
-\ve2\bigg(1-\frac{4(1-6\xi)}{(1-3w_{m})(2-3w_{m})}\bigg)z(\lambda_{3b}^{*})
\Big(x_{3b}^{\text{ini}}+z'(\lambda_{3b}^{*})\big(\lambda_{3b}^{\text{ini}}-\lambda_{3b}^{*}\big)\Big)\big(\exp{(l_{1}\tau)}-1\big)-
\\ & &
+\ve8\frac{1-6\xi}{2-3w_{m}}z(\lambda_{3b}^{*})
\Big(x_{3b}^{\text{ini}}-(1-3w_{m})z'(\lambda_{3b}^{*})\big(\lambda_{3b}^{\text{ini}}-\lambda_{3b}^{*}\big)\Big)\big(\exp{(l_{3}\tau)}-1\big)
\end{array}\right.
\end{equation}
The linearised solutions (\ref{eq:lin3b}) are valid up to the Lyapunov
characteristic time $\tau_{\text{end}}=\frac{1}{l_{2}}=\frac{1}{12\xi}$ and inserting
it in the latter equations we can obtain maximal values of the scale factor and
the Hubble's function valid in the linear approximation.

\begin{figure}
\begin{center}
\epsfig{file=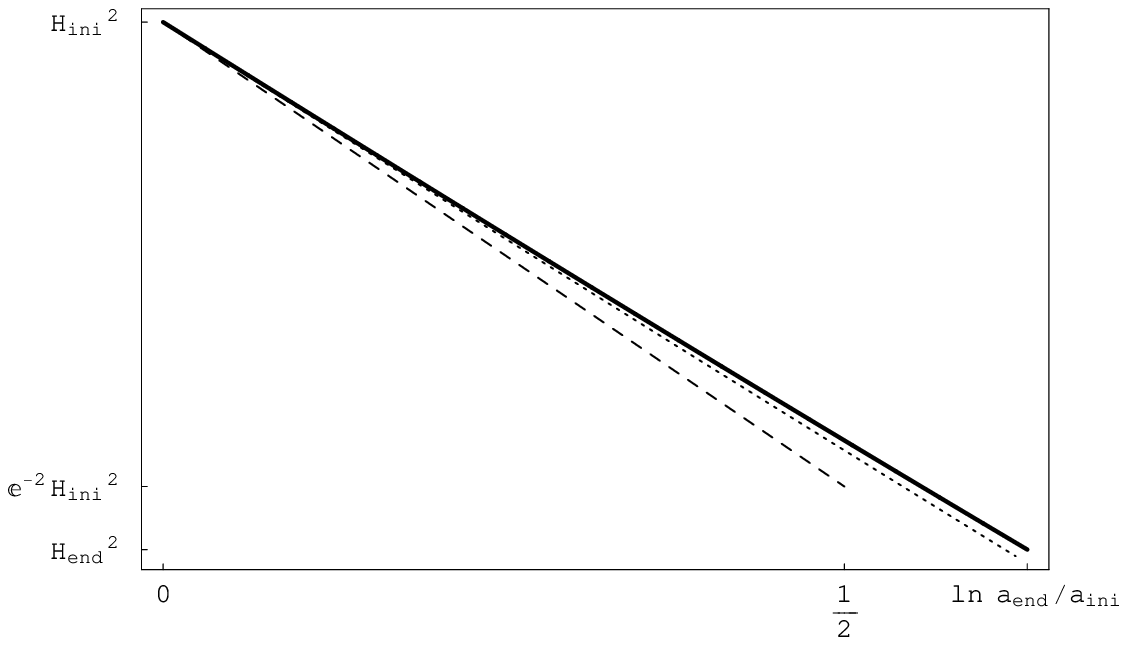,scale=0.65}
\epsfig{file=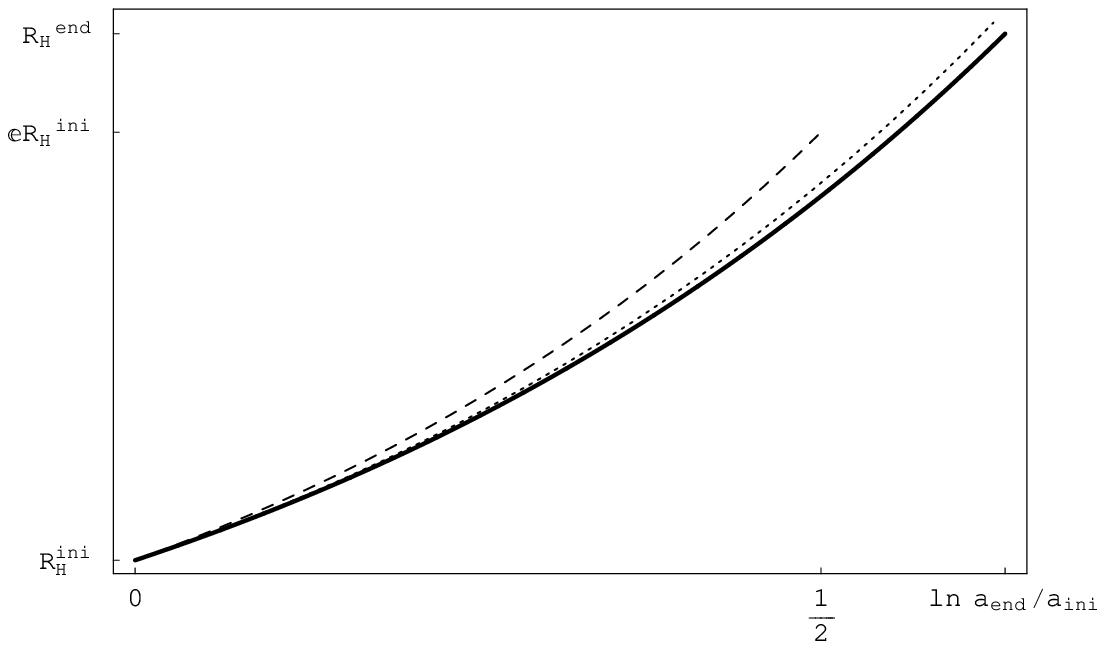,scale=0.65}
\end{center}
\caption{Evolution of $\ln{H^{2}}$ (left panel) and $R_{H}$ (right panel) for a
sample trajectory with $\ve=+1$, $\xi=\frac{1}{16}$,
$z'(\lambda^{*}_{3b})=\frac{1}{\alpha}=100$
in the vicinity of the critical point representing the radiation dominated epoch
for the canonical scalar field. The solid black line corresponds to the
linearised solution (\ref{eq:H2lin3b}), the dashed line corresponds to exact
radiation dominated expansion $\ln{H^{2}}\propto-4\ln{a}$ and the dotted line
corresponds to the numerical solution of the system (\ref{eq:dynsys}).}
\label{fig:lin3b}
\end{figure}

The zero-order approximation ($x^{\text{ini}}_{3b}=0$,
$\lambda^{\text{ini}}_{3b}=\lambda^{*}_{3b}$, $y^{\text{ini}}_{3b}\ne0$ but
$(y^{\text{ini}}_{3b})^{2}\approx0$) is
\begin{equation}
\left\{\begin{array}{lcl}
\ln{\left(\frac{a}{a_{3b}^{\text{ini}}}\right)} & = & 6\xi\tau , \\
\ln{\left(\frac{H}{H_{3b}^{\text{ini}}}\right)^{2}} & = & -24\xi\tau
\end{array}\right.
\end{equation}
and it can be combine to
$$
H^{2}=(H_{3b}^{\text{ini}})^{2}
\left(\frac{a}{{a_{3b}^{\text{ini}}}}\right)^{-4},
$$
which is the exact behaviour of Hubble's function during the radiation
domination era. This approximation is also valid up to
$\tau_{\text{end}}=\frac{1}{12\xi}$ so one can calculate that during radiation
domination epoch the scale factor grows at least
$$
a^{\text{end}}_{3b}=a^{\text{ini}}_{3b}\sqrt{e}.
$$
In figure \ref{fig:lin3b} we present evolution of $\ln{H^{2}}$ and $R_{H}$
as a function of $\ln{a}$ in the vicinity of the critical point representing the
radiation domination era.

\subsection{Matter domination}

The next critical point is located at
\begin{equation}
\label{eq:cp4}
x^{*}_{4} = 0, \quad y^{*}_{4} = 0, \quad \lambda^{*}_{4} \colon z(\lambda)=0
\end{equation}
and $w_{\rm{eff}}$ given by (\ref{eq:weff}) calculated at this point is
$$
w_{\text{eff}} = w_{m}.
$$
We identify this critical point as representing the universe which dynamics is
dominated by the barotropic matter included in the model with the equation of
state parameter $w_{m}$. 

The linearised solutions in the vicinity of this critical point are in the form
\begin{equation}
\begin{array}{lcl}
x_{4}(\tau) & = & \frac{l_{1}}{l_{1}-l_{3}} \bigg(x_{4}^{\text{ini}} - 
z'(\lambda^{*}_{4}) l_{3} \big(\lambda^{\text{ini}}_{4}-\lambda^{*}_{4}\big)\bigg)\exp{(l_{1}\tau)}
- \\
& & - \frac{l_{3}}{l_{1}-l_{3}}\bigg(x_{4}^{\text{ini}} -
z'(\lambda^{*}_{4}) l_{1} \big(\lambda^{\text{ini}}_{4}-\lambda^{*}_{4}\big)\bigg)\exp{(l_{3}\tau)},
\\
y_{4}(\tau) & = & y^{\text{ini}}_{4} \exp{(l_{2}\tau)}, \\
\lambda_{4}(\tau) & = & \lambda^{*}_{4} + \frac{1}{z'(\lambda^{*}_{4})\big(l_{1}-l_{3}\big)}
\bigg(x_{4}^{\text{ini}} - 
z'(\lambda^{*}_{4}) l_{3} \big(\lambda^{\text{ini}}_{4}-\lambda^{*}_{4}\big)\bigg)\exp{(l_{1}\tau)}
- \\
& & - \frac{1}{z'(\lambda^{*}_{4})\big(l_{1}-l_{3}\big)} \bigg(x_{4}^{\text{ini}} - 
z'(\lambda^{*}_{4}) l_{1} \big(\lambda^{\text{ini}}_{4}-\lambda^{*}_{4}\big)\bigg)\exp{(l_{3}\tau)}.
\end{array}
\label{eq:lin4}
\end{equation}
where
\begin{eqnarray}
l_{1} & = &
-\frac{3}{4}\bigg((1-w_{m})+\sqrt{(1-w_{m})^{2}-\frac{16}{3}\xi(1-3w_{m})}\bigg),
\nonumber \\
l_{2} & = & \frac{3}{2}(1+w_{m}), \nonumber \\
l_{3} & = &
-\frac{3}{4}\bigg((1-w_{m})-\sqrt{(1-w_{m})^{2}-\frac{16}{3}\xi(1-3w_{m})}\bigg).
\nonumber
\end{eqnarray}
We need to note that this critical point can became degenerate for two specific
values of $w_{m}$, namely, for $w_{m}=-1$ the second eigenvalue vanishes and
for $w_{m}=\frac{1}{3}$ the third eigenvalue vanish, for any value
of the coupling constant $\xi$, which makes the system in the vicinity of this
critical point degenerated.

From linearised solution (\ref{eq:lin4}) we have
$$
\begin{array}{c}
x_{4}(\tau)^{2}\approx0, \quad y_{4}(\tau)^{2}\approx0, \quad
z\big(\lambda_{4}(\tau)\big)^{2}\approx0
\end{array}
$$
and from (\ref{eq:timerep}) and (\ref{eq:H2}) parametric equations for evolution
of the scale factor and Hubble's function are
\begin{equation}
\left\{\begin{array}{lcl}
\ln{\left(\frac{a}{a_{4}^{\text{ini}}}\right)} & = & \tau, \\
\ln{\left(\frac{H}{H_{4}^{\text{ini}}}\right)^{2}} & = & -3(1+w_{m})\tau.
\end{array}\right.
\label{eq:H2lin4}
\end{equation}
Combining these two expressions we get the Hubble's function as function of the
scale factor during the barotropic matter domination epoch
$$
H^{2} = (H_{4}^{\text{ini}})^{2}
\left(\frac{a}{a_{4}^{\text{ini}}}\right)^{-3(1+w_{m})}.
$$
The linearised solutions (\ref{eq:lin4}) are valid up to the Lyapunov time
$\tau_{\text{end}}=\frac{1}{l_{2}}=\frac{2}{3(1+w_{m})}$,

\begin{equation}
\left\{\begin{array}{lcl}
\ln{\left(\frac{a_4^{\text{end}}}{a_{4}^{\text{ini}}}\right)} & = & \frac{2}{3(1+w_{m})}, \\
\ln{\left(\frac{H_{4}^{\text{end}}}{H_{4}^{\text{ini}}}\right)^{2}} & = & -2.
\end{array}\right.
\end{equation}
One can notice that for dust matter $w_{m}=0$ during the matter domination epoch
the scale factor at least grows
$\frac{a_4^{\text{end}}}{a_{4}^{\text{ini}}} = e^{\frac{2}{3}}\approx1.948$
times.

\begin{figure}
\begin{center}
\epsfig{file=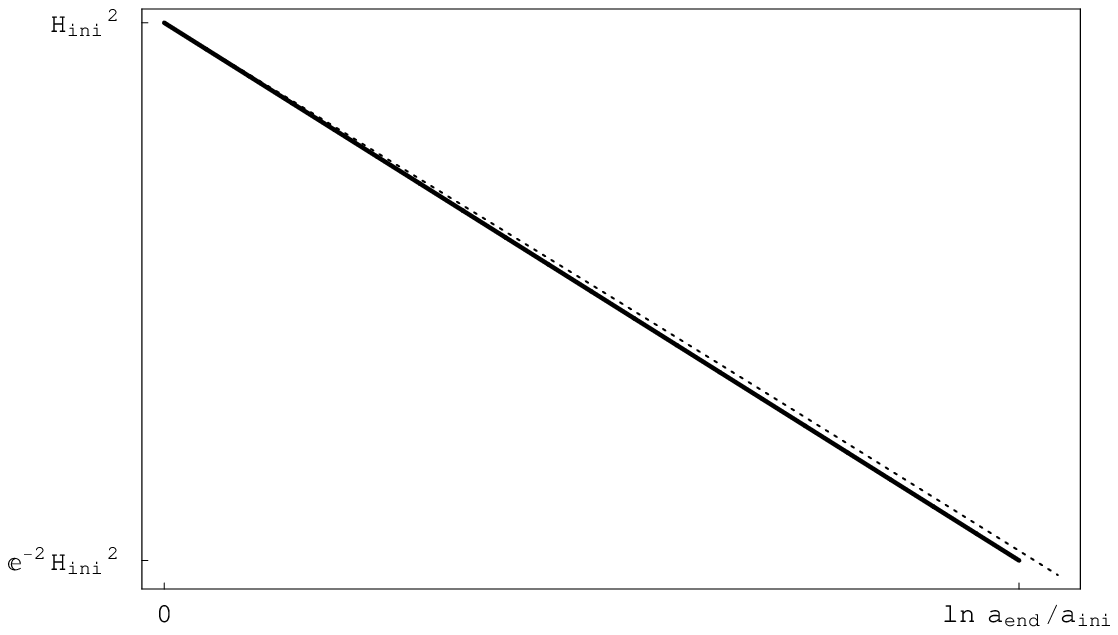,scale=0.65}
\epsfig{file=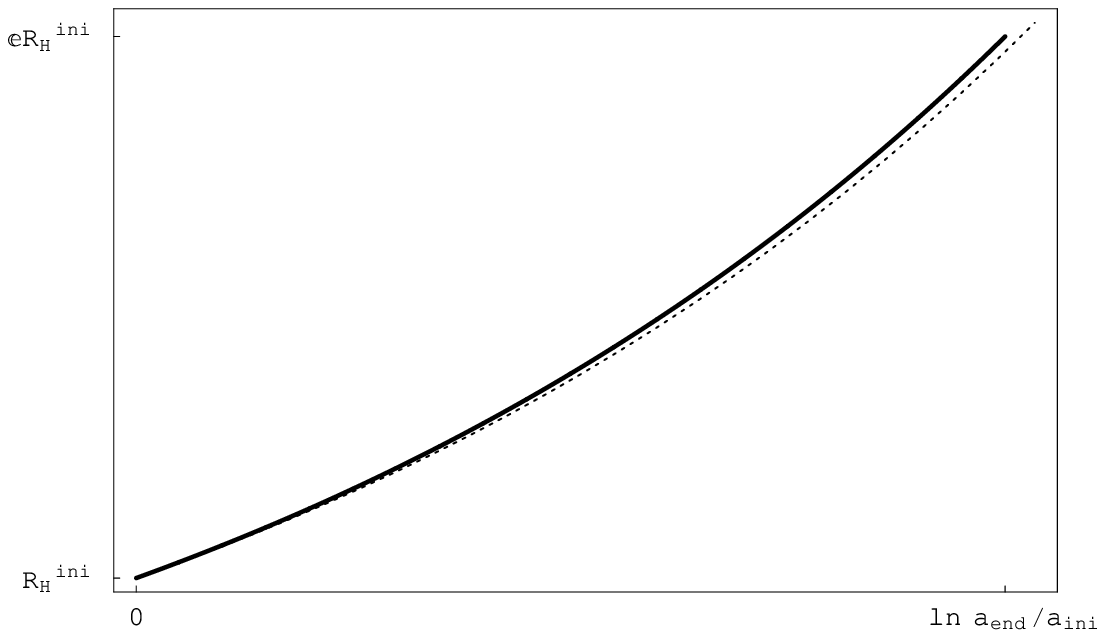,scale=0.65}
\end{center}
\caption{Evolution of $\ln{H^{2}}$ (left panel) and $R_{H}$ (right panel) for a
sample trajectory with $\ve=-1$, $\xi=\frac{1}{2}$,
$z'(\lambda^{*}_{1})=\frac{1}{\alpha}=100$ in
the vicinity of the critical point corresponding to the barotropic matter
dominated universe. The solid black line represents the linear approximation
(\ref{eq:H2lin4}) and the dotted line represents the numerical solution of the
system (\ref{eq:dynsys}).}
\label{fig:lin4}
\end{figure}

In figure \ref{fig:lin4} we present evolution of the scale factor and
Hubble's function in the vicinity of the critical point representing the
barotropic matter domination epoch.

\subsection{The present accelerated expansion epoch}

Finally we proceed to the last critical points located at
\begin{equation}
\label{eq:cp5}
x^{*}_{5} = 0, \quad (y^{*}_{5})^{2} = 1 - \ve6\xi z(\lambda^{*}_{5})^{2} ,
\quad \lambda^{*}_{5} \colon \lambda z(\lambda)^{2} + 4 z(\lambda)
-\frac{\lambda}{\ve6\xi} = 0
\end{equation}
with
$$
w_{\text{eff}} = -1.
$$
There can be more than one such critical points because of the third equation
in (\ref{eq:cp5}) which can have more than one solution. In what follows we will
show that at least one of them represents a stable critical point.

In this case the characteristic equation for eigenvalues of the linearization
matrix calculated at this critical point is in the form
$$
 l^{3} + p l^{2} + q l + r =0
$$
where
$$
\begin{array}{ccl}
p & = & 3(2+w_{m})\big(1-\ve6\xi(1-6\xi)z(\lambda^{*}_{5})^{2}\big),\\
q & = & \big(1-\ve6\xi(1-6\xi)z(\lambda^{*}_{5})^{2}\big)
\Big[-\ve\frac{1}{2}\frac{(y^{*}_{5})^{4}}{z'(\lambda^{*}_{5})} +
12\xi(1+\ve6\xi z(\lambda^{*}_{5})^{2}) +
9(1+w_{m})\big(1-\ve6\xi(1-6\xi)z(\lambda^{*}_{5})^{2}\big)\Big],
\\
r & = & 3(1+w_{m})\big(1-\ve6\xi(1-6\xi)z(\lambda^{*}_{5})^{2}\big)^{2}
\Big[- \ve\frac{1}{2}\frac{(y^{*}_{5})^{4}}{z'(\lambda^{*}_{5})}
+12\xi\big(1+\ve6\xi z(\lambda^{*}_{5})^{2}\big) \Big].
\end{array}
$$

In the most general case without assuming any specific form of the potential
function we are unable to solve this equation. In spite of this we are able to
formulate general conditions for stability of this critical point. This requires
that the real parts of the eigenvalues must be negative. From the Routh-Hurwitz
test \cite{Wiggins:2003} we have that the following conditions should be
fulfilled to assure stability of this critical point
$$
p>0 \qquad \textrm{and} \qquad r > 0 \qquad \textrm{and} \qquad q-\frac{r}{p}>0
$$
Simple inspection of these conditions gives us that, for any matter with
$w_{m}>-2$, $p$ is always positive because of $\big(1-\ve6\xi(1-6\xi)
z(\lambda^{*}_{5})^{2}\big)>0$ due to time transformation and that if $r$ is a
positive quantity it follows that $q-\frac{r}{p}$ is positive too. We conclude
that if the following condition is fulfilled at the critical point
\begin{equation}
Re{[l_{1,2,3}]}<0 \iff - \ve\frac{1}{2}\frac{(y^{*}_{5})^{4}}{z'(\lambda^{*}_{5})}
+12\xi\big(1+\ve6\xi z(\lambda^{*}_{5})^{2}\big) >0
\label{eq:stabcon1}
\end{equation}
it represents a stable critical point with the negative real parts of the
eigenvalues.

In order to simplify this condition let us introduce the following function
\begin{equation}
h(\lambda)= \lambda z(\lambda)^{2} +4 z(\lambda) -\frac{\lambda}{\ve6\xi},
\label{eq:hfun}
\end{equation}
where location of the critical point is the solution to the equation $h(\lambda)=0$,
and obviously $h(\lambda^{*}_{5})=0$.

Let us assume that $\lambda^{*}_{5}\ne0$ it follows from (\ref{eq:hfun}) that
also $z(\lambda^{*}_{5})\ne0$ but $h(\lambda^{*}_{5})=0$. Differentiation of
Eq.~(\ref{eq:hfun}) gives
$$
h'(\lambda^{*}_{5}) = z(\lambda^{*}_{5})^{2} - \frac{1}{\ve6\xi} + 2
z'(\lambda^{*}_{5})\big(\lambda^{*}_{5} z(\lambda^{*}_{5})+2\big)
$$
which after little algebra can be transformed in to the following form
$$
h'(\lambda^{*}_{5}) = - \frac{(y^{*}_{5})^{2}}{\ve6\xi} + 4
\frac{z'(\lambda^{*}_{5})}{(y^{*}_{5})^{2}} \Big(1+\ve6\xi z(\lambda^{*}_{5})^{2}\Big) 
$$
and finally we arrive to the reformulated stability condition~(\ref{eq:stabcon1})
in the form
\begin{equation}
Re{[l_{1,2,3}]}<0 \iff 3\xi\frac{h'(\lambda^{*}_{5})}{z'(\lambda^{*}_{5})} (y^{*}_{5})^{2} >0
\label{eq:stabcon2}
\end{equation}
We have reduced analysis of stability of the critical point representing
accelerated expansion to the simple analysis of the sign of the quantity given
by relation~(\ref{eq:stabcon2}). If we assume that the function $z(\lambda)$ is
a monotonic one, i.e. it is a growing or decreasing function in the interesting region
of the phase space then it follows that if we have at least two critical points
given by (\ref{eq:cp5}) one of them is definitely a stable critical point.

To present the evolution of Hubble's function in the vicinity of this
critical point, as an example, we choose the simple form of $z(\lambda)=
\frac{\lambda}{\alpha}$ function, and values of the parameters $\xi$ and $\alpha$
in range for which there exists only one critical point corresponding the present
accelerated expansion of the universe \cite{Hrycyna:2009zj}. The linearised
solutions in the vicinity of the critical point located at $x_{5}^{*}=0$,
$(y_{5}^{*})^{2}=1$, $\lambda_{5}^{*}=0$ are 

\begin{equation}
\begin{array}{ccl}
x_{5}(\tau) & = & \frac{1}{2\sqrt{\Delta_{5}}} 
\bigg\{
(3+\sqrt{\Delta_{5}})
\Big[x_{5}^{\text{ini}} +
\frac{1}{2\alpha}(3-\sqrt{\Delta_{5}})\lambda_{5}^{\text{ini}}\Big] \exp{(l_{1}\tau)}
- \\ & & \qquad 
-(3-\sqrt{\Delta_{5}})
\Big[x_{5}^{\text{ini}} +
\frac{1}{2\alpha}(3+\sqrt{\Delta_{5}})\lambda_{5}^{\text{ini}}\Big] \exp{(l_{3}\tau)}
\bigg\}, \\
y_{5}(\tau) & = & y_{5}^{*} + (y_{5}^{\text{ini}}-y_{5}^{*}) \exp{(l_{2}\tau)} , \\
\lambda_{5}(\tau) & = & -\frac{\alpha}{\sqrt{\Delta_{5}}}
\bigg\{
\Big[x_{5}^{\text{ini}}+\frac{1}{2\alpha}(3-\sqrt{\Delta_{5}})\lambda_{5}^{\text{ini}}\Big]
\exp{(l_{1}\tau)}- \\ & & \qquad - 
\Big[x_{5}^{\text{ini}}+\frac{1}{2\alpha}(3+\sqrt{\Delta_{5}})\lambda_{5}^{\text{ini}}\Big]
\exp{(l_{3}\tau)}
\bigg\},
\end{array}
\end{equation}
where $l_{1,3}=-\frac{1}{2}\big(3\pm\sqrt{9+\ve2\alpha-48\xi}\big)$ and
$l_{2}=-3(1+w_{m})$ are eigenvalues of the linearization matrix and
$\Delta_{5}=9+\ve2\alpha-48\xi$.

Then keeping only linear terms in initial conditions
$$
\begin{array}{c}
x_{5}(\tau)^{2}\approx0, \quad
z\big(\lambda_{5}(\tau)\big)^{2}\approx0, \quad
\lambda_{5}(\tau)z\big(\lambda_{5}(\tau)\big)\approx0, \\
\\
y_{5}(\tau)^{2} \approx (y_{5}^{*})^{2} +
2y_{5}^{*}\big(y_{5}^{\text{ini}}-y_{5}^{*}\big)\exp{(l_{2}\tau)},
\end{array}
$$
from (\ref{eq:timerep}) and (\ref{eq:H2}) we receive the parametric equations of
evolution of the scale factor and Hubble's function
\begin{equation}
\label{eq:linfin}
\left\{\begin{array}{lcl}
\ln{\left(\frac{a}{a_{5}^{\text{ini}}}\right)} & = & \tau , \\
\ln{\left(\frac{H}{H_{5}^{\text{ini}}}\right)^{2}} & = & 2
y_{5}^{*}\big(y_{5}^{\text{ini}}-y_{5}^{*}\big)\Big(1-\exp{\big(-3(1+w_{m})\tau\big)}\Big)
\end{array}\right.
\end{equation}
Combining these two expressions we get Hubble's function as a function of the
scale factor in the vicinity of the critical point corresponding to the present
accelerated expansion of the universe
$$
H^{2}=(H_{5}^{\text{ini}})^{2}\exp{\left\{2y_{5}^{*}\big(y_{5}^{\text{ini}}-y_{5}^{*}\big)\bigg(1-\Big(\frac{a}{a_{5}^{\text{ini}}}\Big)^{-3(1+w_{m})}\bigg)\right\}}.
$$
One can notice that taking the following limit
$$
H_{\rm fin}^{2}=\lim_{\tau\to\infty}H^{2} = \lim_{a\to\infty}H^{2} =
(H_{5}^{\text{ini}})^{2}\exp{\left\{2y_{5}^{*}\big(y_{5}^{\text{ini}}-y_{5}^{*}\big)\right\}}
\approx (H_{5}^{\text{ini}})^{2}\left(1+2y_{5}^{*}\big(y_{5}^{\text{ini}}-y_{5}^{*}\big)\right)
$$
we get the asymptotic de Sitter expansion.

\begin{figure}
\begin{center}
\epsfig{file=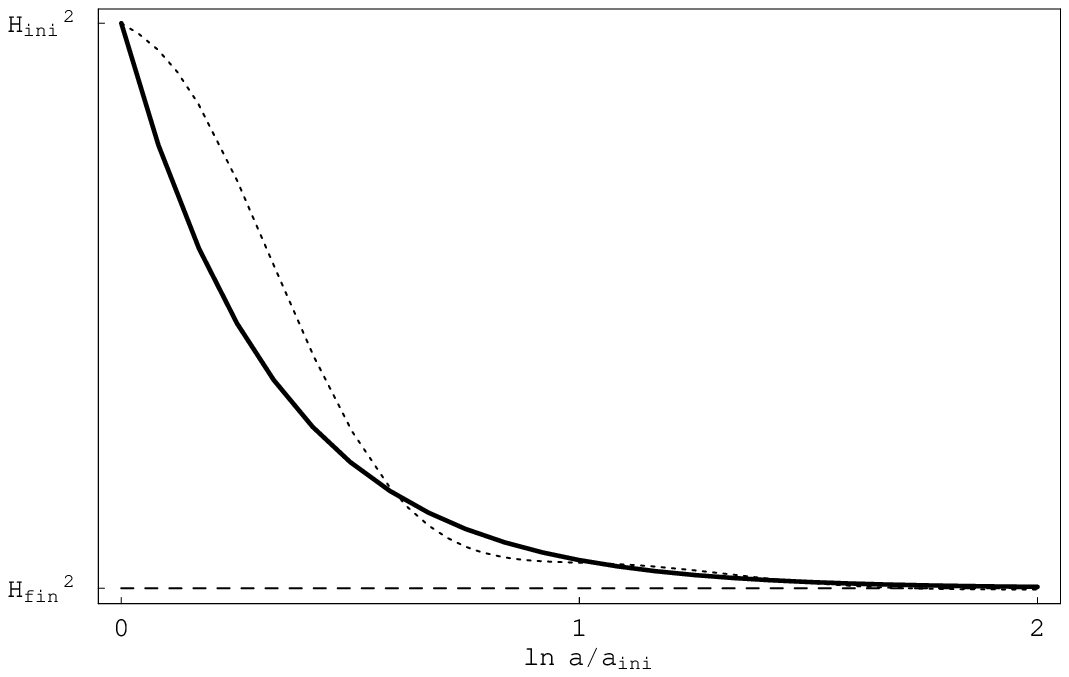,scale=0.65}
\epsfig{file=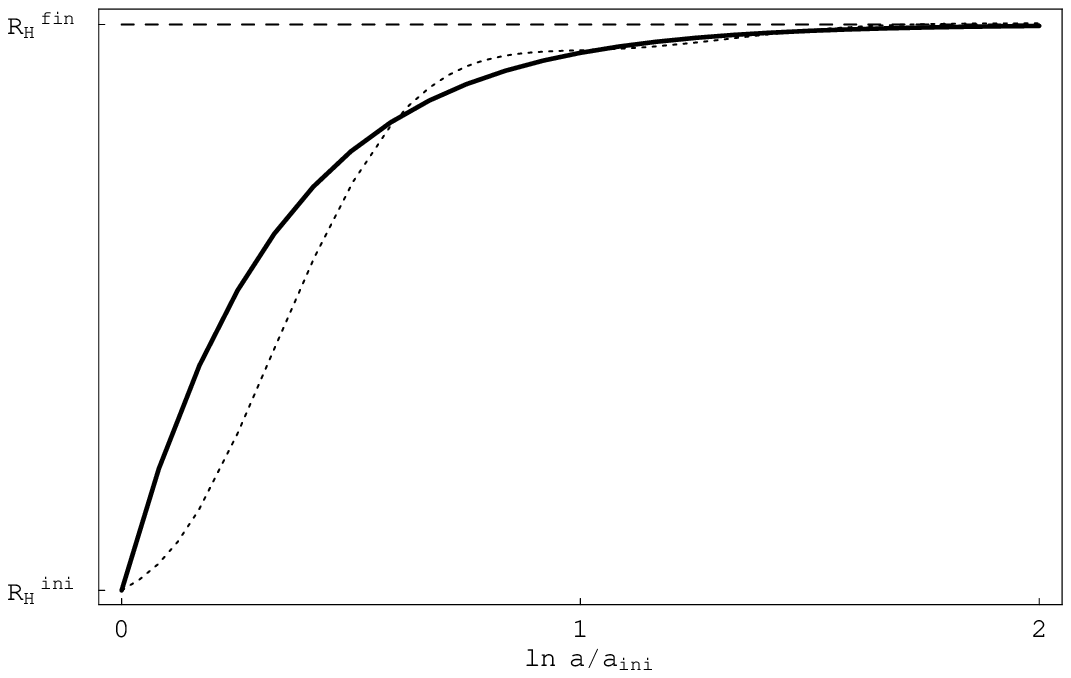,scale=0.65}
\end{center}
\caption{Evolution of $\ln{H^{2}}$ (left panel)
and $R_{H}$ (right panel) as a function of the natural logarithm of the scale
factor $\ln{a}$ for a sample trajectory with $\ve=-1$, $\xi=1$,
$z'(\lambda^{*}_{5})=\frac{1}{\alpha}=20$,
$y^{\text{ini}}_{5}-y^{*}_{5}=-\frac{1}{100}$ in the vicinity of the critical point
representing the present accelerated expansion of the universe. The solid black line represents the
linear approximation (\ref{eq:linfin}) and the dotted line corresponds to
numerical solution of the dynamical system (\ref{eq:dynsys}).}
\label{fig:lin5}
\end{figure}

In figure \ref{fig:lin5} we present the evolution of $\ln{H^{2}}$ and $R_{H}$ as
a function of $\ln{a}$ in the vicinity of this critical point.

\section{Summary and Conclusions}

Modern cosmology becomes very similar to the particle physics. Both theories
have parameters and characteristic energetic cut offs. They are the effective
description of deeper physics which is currently unknown. The values of these
parameters should be obtained form more fundamental theories or from
observations. In cosmology ($\Lambda$CDM model is called Standard Cosmological
Model) the role of such a parameter plays the cosmological constant. Our
proposition is to extend this paradigm in which matter content is described in
terms of barotropic perfect fluid by introduction additional scalar field
non-minimally coupled to gravity. As a result we discover new evolutional path
which open new perspectives of description of cosmological evolution in unified
way. In this scheme the inflation era appears in natural way and it is not put
into the $\Lambda$CDM scenario by hand.

In this paper we have shown that the all important epochs in the evolution of
the universe can
be represented by the critical points of the dynamical system arising from the
non-minimally coupled scalar field cosmology in spite of not assuming a form of the
potential function. We have shown that for the positive coupling constant there
exists a past finite scale factor singularity for both types of the scalar
fields. Additionally all the intermediate states are transient one, i.e. they
are represented by an unstable critical points in the phase space and last for
an finite amount of time. The existence of the radiation dominated era is purely
the result of the evolution of the non-minimally coupled scalar field.

\begin{figure}
\begin{center}
\epsfig{file=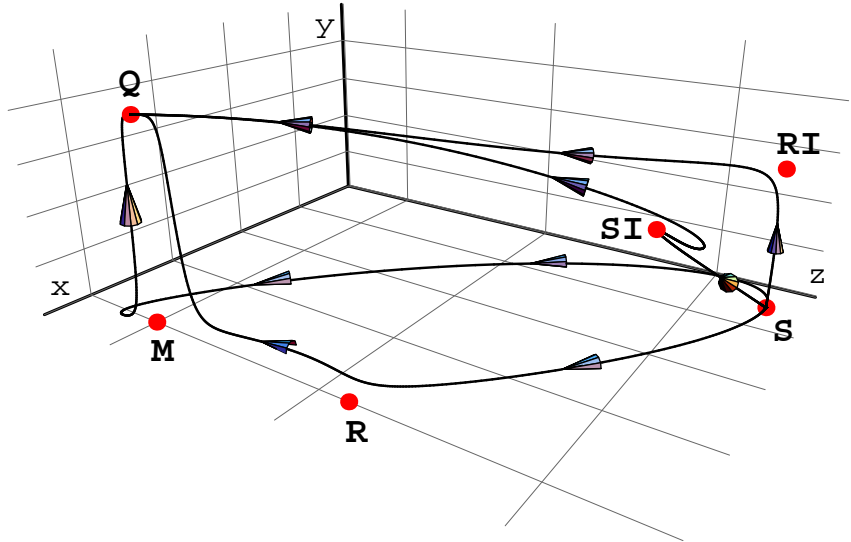,scale=1}
\end{center}
\caption{The phase space portrait for the model with the cosmological constant and
the canonical scalar field ($\ve=+1$) with $\xi=1/8$ and the dust matter
$w_{m}=0$. The critical points are: $S$ -- the finite scale factor singularity,
$RI$ -- the rapid-roll inflation, $SI$ -- the slow-roll inflation, $R$ -- the
radiation dominated era, $M$ -- the barotropic matter dominated era and $Q$ --
the quintessence era. Note that the critical points representing the finite scale
factor singularity, the rapid-roll inflation and the slow-roll inflation have
the same value of coordinate $z$.}
\label{fig:3}
\end{figure}

\begin{figure}
\begin{center}
\epsfig{file=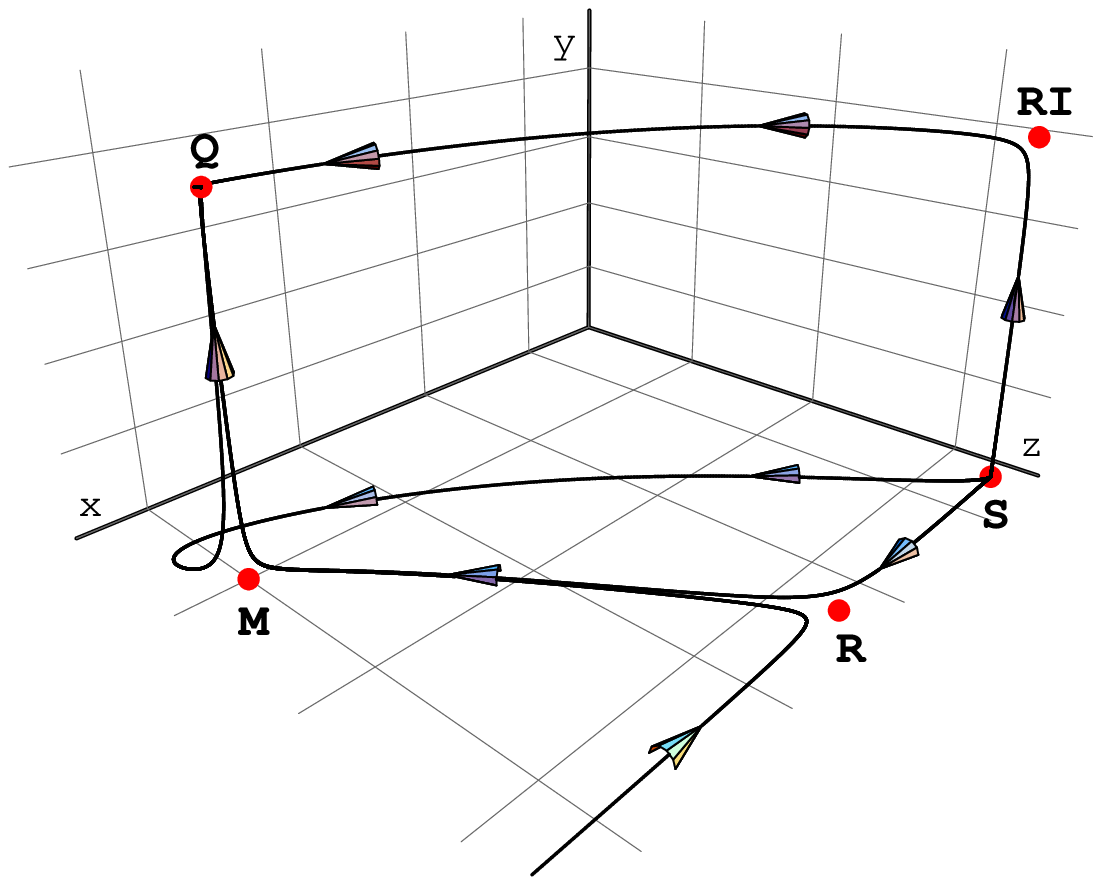,scale=1}
\end{center}
\caption{The phase space portrait for the model with the cosmological constant and
the phantom scalar field ($\ve=-1$) with $\xi=1/4$ and the dust matter
$w_{m}=0$. The critical points are: $S$ -- the finite scale factor singularity,
$RI$ -- the rapid-roll inflation, $R$ -- the radiation dominated era, $M$ -- the
barotropic matter dominated era and $Q$ -- the quintessence era. In the case of
the phantom scalar field the critical point representing slow-roll inflation is
not present. The critical pints denoted as $S$, $RI$ and $R$ have the same value
of coordinate $z$.}
\label{fig:4}
\end{figure}

For the canonical scalar field $\ve=+1$ and $0<\xi<1/6$ we can construct the
unique evolutional path represented by the trajectory in the phase space which
travels in the vicinity of the following critical points (figure \ref{fig:3})
$$
1 \mapsto 2b \mapsto 3b \mapsto 4 \mapsto 5 
$$
and for the phantom scalar field $\ve=-1$ and $\xi>1/6$ (figure \ref{fig:4})
$$
1 \mapsto 2a \mapsto 3a \mapsto 4 \mapsto 5
$$

\begin{figure}
\begin{center}
\epsfig{file=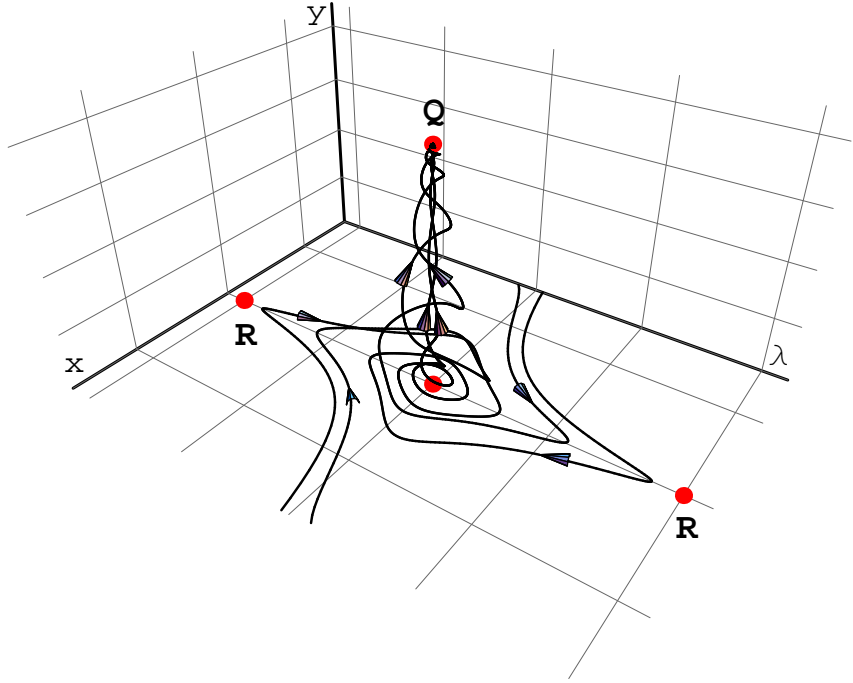,scale=1}
\end{center}
\caption{The phase space portrait representing twister type behaviour.
Trajectories in this type solution interpolate between three major epochs in the
history of universe: $R$ -- the radiation dominated universe with
$w_{\rm{eff}}=\frac{1}{3}$, the matter domination epoch (an unstable focus type
critical point) and $Q$ -- the quintessence domination epoch with $w_{\rm{eff}}=-1$.
This type of evolution does not depend on the form of assumed function $z(\lambda)$
(i.e. the form of the scalar field potential) and is generic for the
canonical scalar field cosmologies ($\ve=+1$) with $\xi>0$ and the barotropic
matter with equation of state parameter $-1<w_{m}<1/3$.}
\label{fig:5}
\end{figure}

Within one framework of non-minimally coupled scalar field cosmology we were
able to unify all the major epochs in the history of the universe (see
figure \ref{fig:5} for twister type behaviour where trajectories interpolate
between radiation era, matter domination era and quintessence epoch). 

From the analysis presented in this paper one can draw the general conclusion
that if the non-minimal coupling constant is present and is different from the
conformal coupling $\xi\ne1/6$ then new evolutional types emerge forming the
structure of the phase space nontrivial and richer. Moreover the coupling
constant gives us the effect of continuation (glues the evolution) between
different cosmological epochs which is very attractive in cosmology, serving as
a potential explanation of the global properties of the universe.

\appendix
\section{The Center Manifold Theorem for three-dimensional dynamical systems}
\label{appa}

For the sake of completeness we present here the theorem concerning behaviour of
nonlinear dynamical system in the vicinity of degenerated critical point.
Expanded discussion can be found, for example, in books by Perko
\cite{Perko:2001} or Wiggins \cite{Wiggins:2003}.

Suppose we consider the following 3-dimensional nonlinear dynamical system
\begin{equation}
\label{appa:1}
\dot{\mathbf{x}} = f(\mathbf{x}), \qquad \mathbf{x}\in \mathbb{R}^{3}
\end{equation}
We are interested in the nature of solution to this dynamical system near fixed
point $\bar{\mathbf{x}}$ for which $f(\bar{\mathbf{x}})=0$.

First, we transform the fixed point $\mathbf{x}=\bar{\mathbf{x}}$ of
(\ref{appa:1}) to the origin using the transformation
$\mathbf{y}=\mathbf{x}-\bar{\mathbf{x}}$. Then the system (\ref{appa:1})
becomes
\begin{equation}
\dot{\mathbf{y}} = f(\bar{\mathbf{x}}+\mathbf{y}), \qquad \mathbf{y}\in
\mathbb{R}^{3}
\end{equation}
then Taylor expansion of $f(\bar{\mathbf{x}}+\mathbf{y})$ about
$\mathbf{x}=\bar{\mathbf{x}}$ gives
\begin{equation}
\label{appa:2}
\dot{\mathbf{y}} = A \mathbf{y} + R(\mathbf{y}),\qquad \mathbf{y}\in
\mathbb{R}^{3}
\end{equation}
where $A=Df(\bar{\mathbf{x}})$ ia a linearization matrix calculated at the fixed
point, $R(\mathbf{y})=\mathcal{O}(|\mathbf{y}|^{n})$ and we have used
$f(\bar{\mathbf{x}})=0$.

From now on we will assume that the linearization matrix has purely real
eigenvalues and one is zero  $l_{1}=0$, one positive $l_{2}>0$ and one negative
$l_{3}<0$. From
elementary linear algebra we can find a linear transformation $P$ which
transforms the linear part of equation~(\ref{appa:2}) into a diagonal form
\begin{equation}
\left(\begin{array}{c} \dot{u} \\ \dot{v} \\ \dot{w} \end{array}\right) =
\left(\begin{array}{ccc} 0 & 0 & 0 \\ 0 & l_{2} & 0 \\ 0 & 0 & l_{3}
\end{array}\right)\left(\begin{array}{c} u \\ v \\ w \end{array}\right)
\end{equation}
with a linear transformation of variables 
$$P^{-1}\mathbf{y} \equiv \left(\begin{array}{c} u \\ v \\ w \end{array}\right)$$
and the matrix $P$ is constructed from the corresponding eigenvectors of the
linearization matrix $A$. Using this same linear transformation to transform the
coordinates of the nonlinear part of the system~(\ref{appa:2}) gives the
following 
\begin{equation}
\label{appa:3}
\begin{array}{ccr}
\dot{u} & = &  F\big(u,v,w\big),\\
\dot{v} & = & l_{2}v + G\big(u,v,w\big),\\
\dot{w} & = & l_{3}w + H\big(u,v,w\big).
\end{array}
\end{equation}
where $F(u,v,w)$, $G(u,v,w)$ and $H(u,v,w)$ are polynomial in the coordinates.
The fixed point $(u,v,w)=(0,0,0)$ is unstable due to the existence of a
1-dimensional unstable manifold associated with the negative eigenvalue
$l_{3}<0$.

\begin{definition}\emph{\bf (Center Manifold)}
An invariant manifold will be called a center manifold for (\ref{appa:3}) if it
can be locally represented by
\begin{equation}
W^{c}(0) = \Big\{(u,v,w)\in\mathbb{R}^{3} \big| v=h_{1}(u) ,w=h_{2}(u), |u|<\delta,
h_{i}(0)=0, h_{i}'(0)=0, i=1,2\Big\}
\end{equation}
for $\delta$ sufficiently small.
\end{definition}

\begin{theorem}\emph{\bf(Existence)}
There exists a $\mathbf{C}^{r}$ center manifold for (\ref{appa:3}). The dynamics
of (\ref{appa:3}) restricted to the center manifold is, for $\eta$ sufficiently
small, given by the following 1-dimensional vector field
\begin{equation}
\dot{\eta} = F\big(\eta,h_{1}(\eta),h_{2}(\eta)\big).
\end{equation}
\end{theorem}

From the fact that the center manifold is invariant under the dynamics
generated by (\ref{appa:3}) we obtain
\begin{equation}
\begin{array}{ccl}
\dot{u} & = & F\big(u,h_{1}(u),h_{2}(u)\big), \\
\dot{v} & = & h_{1}'(u)\dot{u} = l_{2}h_{1}(u) + G\big(u,h_{1}(u),h_{2}(u)\big),\\
\dot{w} & = & h_{2}'(u)\dot{u} = l_{3}h_{2}(u) + H\big(u,h_{1}(u),h_{2}(u)\big),
\end{array}
\end{equation}
which yields the following quasilinear differential equation for $h_{1}(u)$ and
$h_{2}(u)$
\begin{equation}
\label{appa:4}
\begin{array}{c}
\mathcal{N}\big(h_{1}(u)\big)=h_{1}'(u)F\big(u,h_{1}(u),h_{2}(u)\big) -
l_{2}h_{1}(u) - G\big(u,h_{1}(u),h_{2}(u)\big) =0,\\
\mathcal{N}\big(h_{2}(u)\big)=h_{2}'(u)F\big(u,h_{1}(u),h_{2}(u)\big) -
l_{3}h_{2}(u) - H\big(u,h_{1}(u),h_{2}(u)\big) =0,
\end{array}
\end{equation}
and the following theorem (see \rm{Theorem 18.1.4} in Wiggins 
\cite[p. 248]{Wiggins:2003}) justify solving (\ref{appa:4}) approximately via
power series expansion:
\begin{theorem}\emph{\bf(Approximation)}
Let $\phi \colon \mathbb{R}\to\mathbb{R}$ be a $\mathbf{C}^{1}$ mapping with
$\phi(0)=\phi'(0)=0$ such that
$\mathcal{N}\big(\phi(u)\big)=\mathcal{O}\big(|u|^{q}\big)$ as
$u\to 0$ for some $q>1$. Then
$$
|h(u)-\phi(u)| = \mathcal{O}\big(|u|^{q}\big) \qquad {\rm as}\qquad u\to0.
$$
\end{theorem}
This theorem allows us to compute the center manifold to any desired degree of
accuracy by solving (\ref{appa:4}) to the same degree of accuracy.

\bibliographystyle{JHEP}
\bibliography{../darkenergy,../quintessence,../quartessence,../astro,../dynamics,../standard,../inflation,../sm_nmc,../singularities}

\end{document}